\documentclass[12pt]{article}
\usepackage[width=8.50in, height=11.00in, left=1.00in, right=1.00in, top=1.00in, bottom=1.00in]{geometry}
\usepackage{amsmath}
\usepackage{amsfonts}
\usepackage{amssymb}
\usepackage{graphicx}
\usepackage[numbers,sort&compress]{natbib}

\begin{document}
	
\title{\textbf{How Special Is the Solar System?}\footnote{Chapter for the book \textit{Consolidation of Fine Tuning.}}}
\author{\textbf{Mario Livio}\\
Dept.\ of Physics and Astronomy, University of Nevada, Las Vegas,\\
4505 South Maryland Parkway, Las Vegas, NV 89154, USA;\\
Visiting Scholar, Dept.\ of Particle Physics and Astrophysics, Faculty of Physics,\\
The Weizmann Institute of Science, Rehovot 76100, Israel;\\ 
E-mail: dr.mario.livio@gmail.com}
\date{}
\maketitle
\begin{abstract}
	Given the fact that Earth is so far the only place in the Milky Way galaxy known to harbor life, the question arises of whether the solar system is in any way special. To address this question, I compare the solar system to the many recently discovered exoplanetary systems. I identify two main features that appear to distinguish the solar system from the majority of other systems: (i)~the lack of super-Earths, (ii)~the absence of close-in planets. I examine models for the formation of super-Earths, as well as models for the evolution of asteroid belts, the rate of asteroid impacts on Earth, and of snow lines, all of which may have some implications for the emergence and evolution of life on a terrestrial planet.
	
	Finally, I revisit an argument by Brandon Carter on the rarity of intelligent civilizations, and I review a few of the criticisms of this argument.
\end{abstract}
\newpage

\section{Introduction\label{sec1}}
The discovery of thousands of extrasolar planets in the Milky Way galaxy in recent years (e.g., \citep{Batalha2014} and references therein) has led to the realization that our Galaxy may contain (on the average) as many as 0.16 Earth-size planets and 0.12 Super-Earths (planets with a mass of a few Earth masses) per every M-dwarf Habitable Zone~\citep{D+C2015}. The habitable zone is that relatively narrow region in orbital distances from the central star that allows for liquid water to exist on the surface of a rocky planet. Given that it is still the case that to date Earth is the only place known to support complex life (or any life form for that matter), the plethora of potentially habitable extrasolar planets raises the important question of whether the solar system is in any sense special. This question is further motivated by the so-called ``Fermi Paradox''---the absence of any signs for the existence of other intelligent civilizations in the Milky Way~\citep{Jones1985,L+S2017}. It is interesting to explore, therefore, the status of the solar system in the context of the entire known population of extrasolar planets.

The solar system includes eight planets (suggestions for a ninth planet are yet to be confirmed; \citep{B+B2016}), and two belts composed of generally smaller bodies---the asteroid belt and the Kuiper belt. In an attempt to compare the solar system to other exoplanetary systems, \citet{M+L2015} first considered the statistical distributions of orbital separations and eccentricities of the observed exoplanetary orbits. To this goal, and in order to allow for a more meaningful statistical analysis, they performed a transformation on the available data that makes them closer to a normal distribution~\citep{B+C1964}. Specifically, they transformed the data with the function
\begin{equation}
y_{\lambda} (a)=\begin{cases}
\frac{a^{\lambda}-1}{\lambda} &\text{if $\lambda\neq0$}\\
\log a &\text{if $\lambda=0$~~,}
\end{cases}
\end{equation}
where $a$ is the examined parameter (e.g., eccentricity or semi-major axis), and $\lambda$ is a constant determined through the process described below. The maximum liklihood estimator of the mean of the transformed data is
\begin{equation}
\overline{y}_\lambda=\sum^n_{i=1} \frac{y_{\lambda,i}}{n}~~,
\end{equation}
where $y_{\lambda,i}=y_\lambda(a_i)$ and $a_i$ is the $i$th measurement of the total of $n$. Similarly, the maximum likelihood estimator of the variance of the transformed data is
\begin{equation}
s_\lambda^2=\sum^n_{i=1} \frac{(y_{\lambda,i}-\overline{y}_\lambda)^2}{n}~~.
\end{equation}
\citet{M+L2015} chose $\lambda$ such that they maximize the log likelihood function
\begin{equation}
\ell(\lambda)=\frac{n}{2}\log(2\pi)-\frac{n}{2}-\frac{n}{2}\log s^2_\lambda+(\lambda-1)\sum^n_{i-1}\log a_i~~.
\end{equation}
The new distribution, $y_\lambda(a)$, becomes an exact normal distribution if $\lambda=0$ or 1/$\lambda$ is an even integer.

\begin{figure}
	\centering
	\includegraphics[]{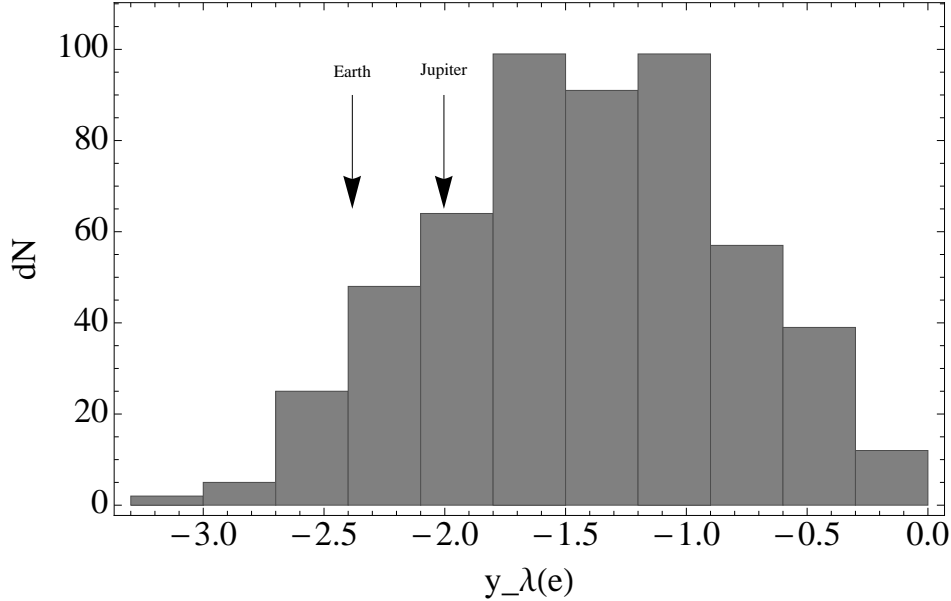}
	\caption{Box--Cox transformed distribution of exoplanet eccentricities. The total number of exoplanets is 539.\label{f1}}
\end{figure}

Figure~\ref{f1} shows the Box--Cox transformed eccentricities for 539 extrasolar planets with measured eccentricities. As we can see, Jupiter lies at $-0.97\sigma$ from the mean, and the Earth (the unlabeled arrow) is at $-1.60\sigma$. In other words, while the eccentricities in the solar system (ranging from $e=0.0068$ for Venus to $e=0.21$ for Mercury) are on the low side compared to the general distribution of exoplanets, they are not altogether exceptional. Furthermore, since the mean eccentricity of a planetary system appears to be anticorrelated with the number of the planets in the system \citep[e.g.,][]{L+T2015}, the relatively low mean eccentricity of the solar system is actually the one expected for an eight-planet system (a conclusion that is further strengthened when selection biases are taken into account,  \citep[e.g.,][]{Moorheadetal2011,Hoggetal2010,Zakamskaetal2011}).

\begin{figure}
	\centering	
	\includegraphics[]{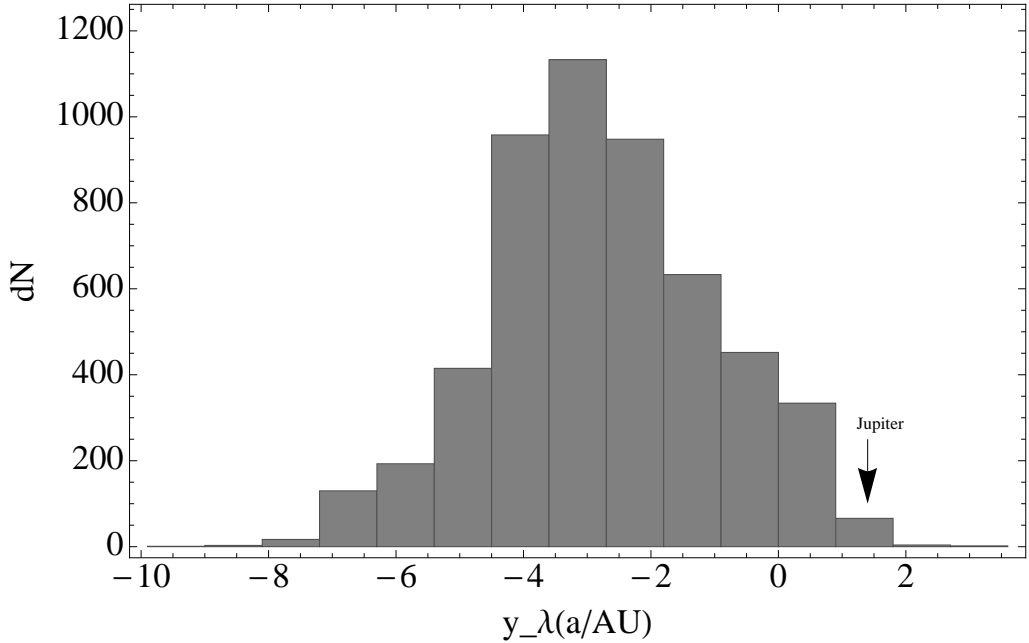}
	\caption{Box--Cox transformed distribution of exoplanet semi-major axis, including all planet candidates. The total number of planets is 5289.\label{f2}}
\end{figure}

An examination of the transformed distribution of the semi-major axis for 5289 candidate planets in the Kepler sample, shows that Jupiter lies at $2.4\sigma$ from the mean (Fig.~\ref{f2}). At first blush, this may suggest that the largest planet in the solar system is rather special, but a closer inspection reveals that this fact is most likely resulting from selection effects. For example, if we repeat the analysis after removing planets found through transits (a method favoring planets that are close in, which is only complete in the Kepler sample for periods of up to one year, \citep{W+F2015}), we find that Jupiter's deviation from the mean is reduced to $1.44\sigma$. This trend appears to be further strengthened by the fact that the number of detections by direct imaging is constantly increasing, suggesting that there may indeed exist an  entire population of planets with semi-major axes longer than those of Jupiter, which have so far escaped detection.

\begin{figure}
	\centering
	\includegraphics[]{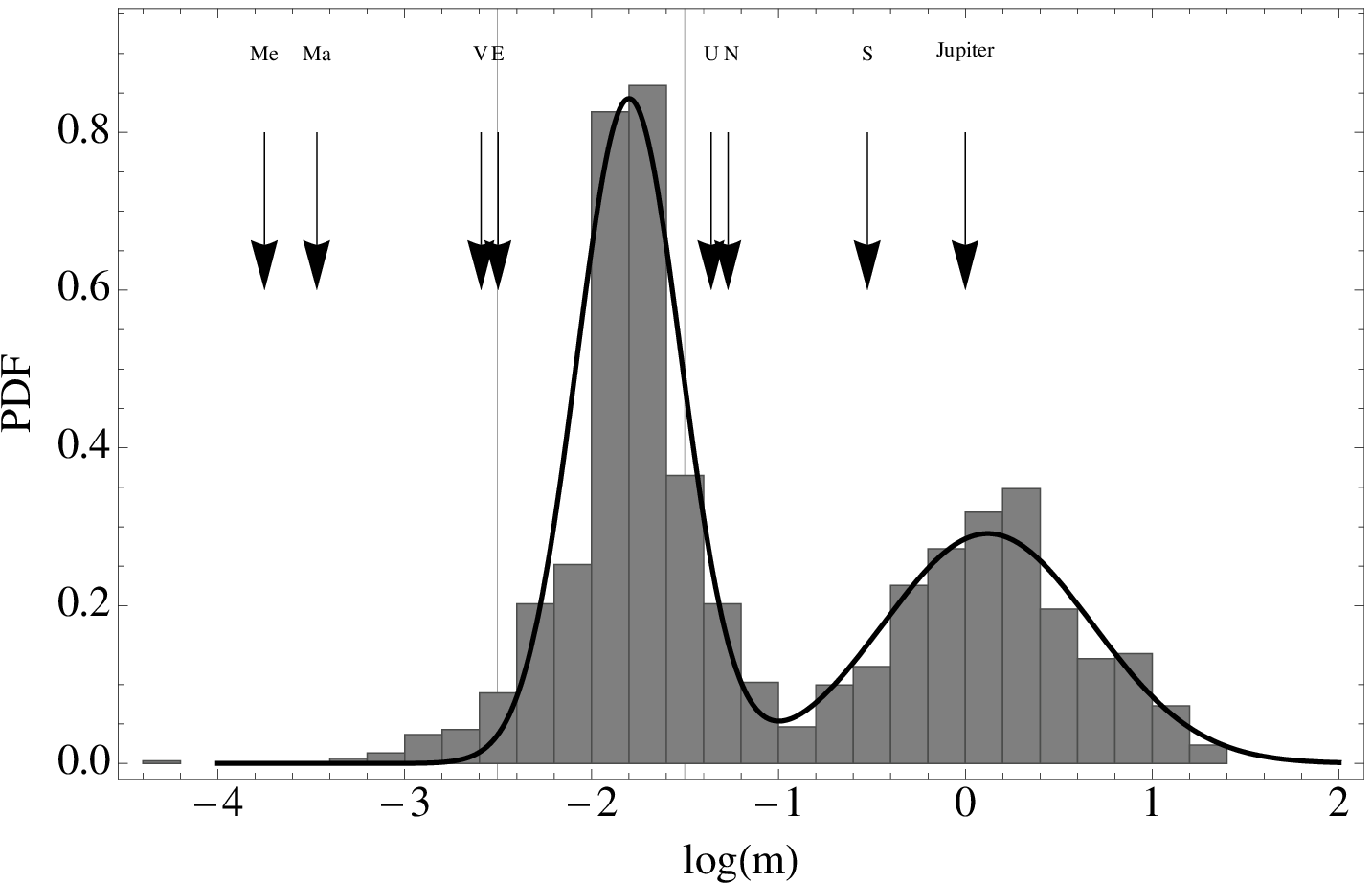}
	\caption{The exoplanet mass distribution. The arrows indicate the masses of the planets in the solar system. The vertical lines show the range of planets considered to be super-Earths. The total number of exoplanets is 1516.\label{f3}}
	\centering
	\includegraphics[]{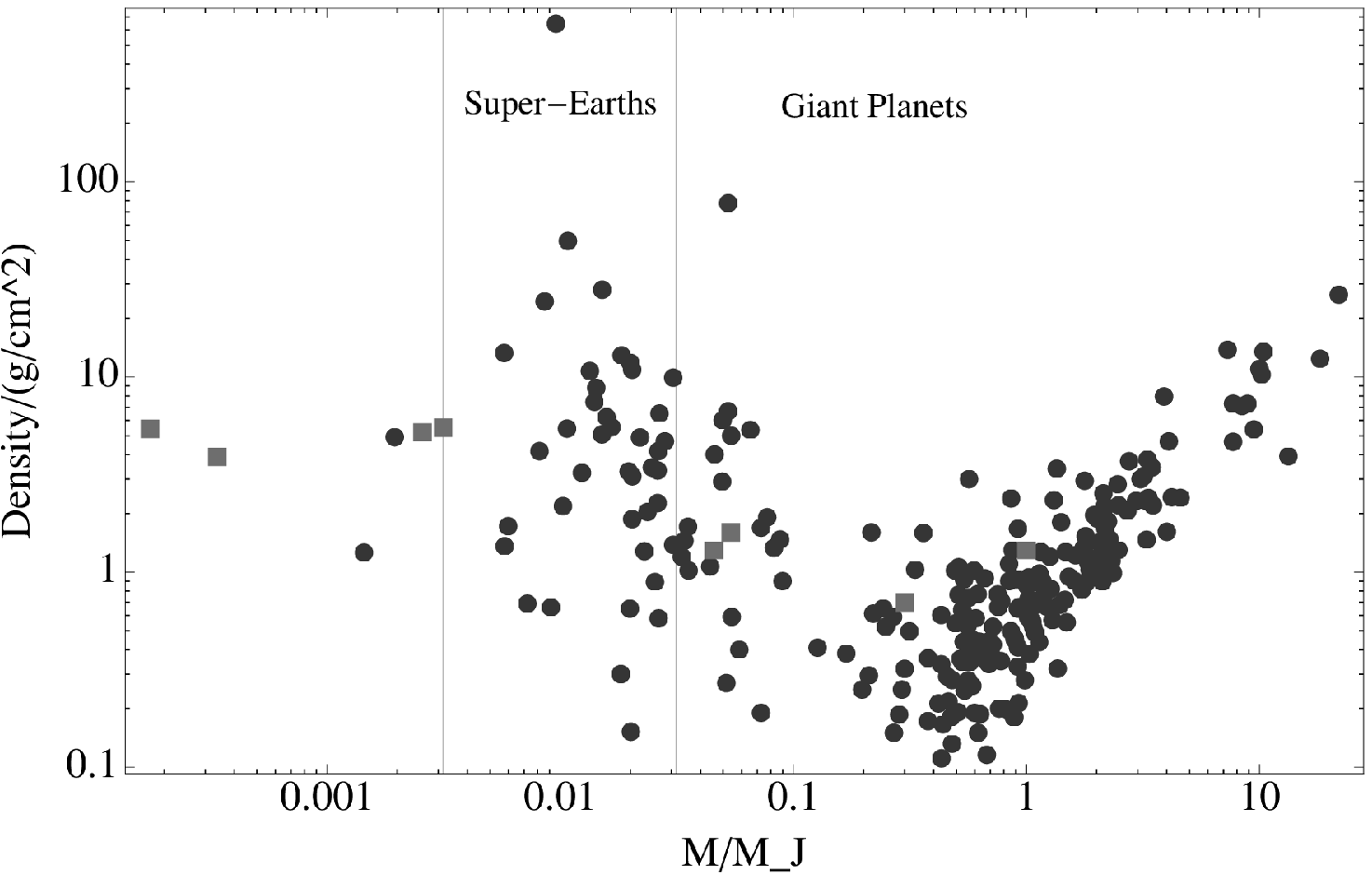}
	\caption{Densities of planets as a function of their mass. The dark circles show the exoplanets (a total of 287) and the lighter gray squares the planets in our solar system. The vertical lines again indicate the range of super-Earths.\label{f4}}
\end{figure}

\citet{M+L2015} also found that the masses of the gas giants in the solar system fit nicely within the distribution of extrasolar planets (Fig.~\ref{f3}; the observations of small terrestrial planets are most likely still affected by selection effects), as do their densities (Fig.~\ref{f4}). \citet{Dressingetal2015} showed, in addition, that the Earth and Venus can be modeled with a ratio of iron to magnesium silicate similar to that of the low-mass extrasolar planets.

Is there anything, then, in terms of planetary and orbital properties, that makes the solar system (even somewhat) special? \citet{M+L2015} identified two properties that are at least intriguing: (i)~The lack of super-Earths, and (ii)~The lack of very close-in planets. Let's briefly discuss each one of these characteristics.

\section{Lack of Super-Earths\label{sec2}}

Super-Earths are planets with masses typically between 1--10 Earth masses. They are very common in exoplanetary systems. In fact, more than half of the observed Sun-like stars in the solar neighborhood are orbited by one or more super-Earths with periods of days to months  (e.g., \citep{Mayoretal2011,Batalhaetal2013,Fressinetal2013,Burkeetal2015}). Furthermore, systems observed to contain a super-Earth usually have more than one. Most recently, observations of the star Trappist-1 found that it hosts at least seven Earth-size planets. Their orbits, six of which form a near-resonant chain (with orbital periods of 1.51, 2.42, 4.04, 6.06, 9.1, and 12.35 days), suggest that these planets formed farther from their host star and then migrated inwards (e.g., \citep{Tamasoetal2017}). If super-Earths form via mergers of inwardly migrating cores, then having more than one super-Earth is theoretically expected (e.g., \citep{Cossouetal2014}).

The fact that the solar system does not contain any close-in super-Earths does set it somewhat apart from most observed exoplanetary systems. Before I examine potential \textit{reasons} for this lack of super-Earths, it is interesting to contemplate on whether not containing super-Earths (by a planetary system) can in any way be related to the emergence of life in such a system. It is interesting to note that the presence of super-Earths may effect the formation process and properties of terrestrial planets. In particular, \citet{Izidoroetal2014} found that if a super-Earth migrates slowly enough through the habitable zone of its host star (the range of orbital radii that allows for liquid water to exist on a rocky planetary surface), then terrestrial planets subsequently forming in this region would be rich in volatiles---far from being an Earth twin. A super-Earth at a very small orbital radius could also (in principle at least) disturb the dynamical stability of a terrestrial planet in the habitable zone. In contrast, the terrestrial planets in the solar system are expected to remain dynamically stable until the Sun expands to become a red giant (e.g., \citep{Laughlin2009}). Finally, a super-Earth relatively close to the orbital distance of a terrestrial planet could significantly affect the rate of asteroid impacts on such a planet (see Section~\ref{sec53} for an extensive discussion).

\section{Lack of Close-In Planets or Debris\label{sec3}}

The second element that appears to distinguish the solar system from most observed exoplanetary systems is the lack of very close-in planets (or, for that matter, any type of mass). Specifically, while Mercury is at 0.39~AU from the Sun, other exoplanetary systems harbor planets much closer to their host star~\citep{B+L2015}. In particular, most systems observed to have three or more planets contain a planet with a semi-major axis smaller than that of Mercury.

Again, it is not clear to what extent the absence of close-in planets is related (if at all) to the Earth's habitability. In the next section I examine the conditions that are necessary for the formation of super-Earths, and their potential implications for the two characteristics that make the solar system somewhat special.

\section{On the Formation of Super-Earths\label{sec4}}

Super-Earths either form \textit{in~situ} with no significant migration through the protoplanetary disk, or they form outside the snow line (the distance from the central star where the temperature is sufficiently low for water to solidify; \citep{Lecaretal2006,M+L2012}), and then migrate inwards. \citet{C+L2013} used observations of super-Earths with orbital periods of $P\lesssim100$ days to construct a minimum-mass extrasolar nebula (MMEN) with a surface density of the gas disk of ($R$ is the distance from the star)
\begin{equation}
\Sigma_\mathrm{MMEN}=9900\left(\frac{R}{1~\mathrm{AU}}\right)^{-1.6}~\mathrm{g~cm}^{-2}~~.
\end{equation}
Equation~(5) gives a somewhat higher value for the surface density than the minimum mass solar nebula (MMSN; which gives 1700~g cm$^{-2}$ at $R=1$~AU, required to form the planets in our solar system~\citep{Weiden1977,Hayashi1981}). However, it is not clear whether the MMSN is applicable at distances smaller than 0.4~AU (inside Mercury's orbit).
\citet{H+M2012,H+M2013} found that to form super-Earths \textit{in~situ} required having about 50--100~$M_{\oplus}$ of rocky material interior to 1~AU. Other researchers suggested that super-Earths form farther out in the disk (where solid material is more readily available) and migrate inwards (e.g., \citep{T+P2007,I+L2010,Cossouetal2014}). 

\citet{M+L2016} showed that in fully turbulent disk models (i.e., disks in which the magneto-rotational instability generates viscosity throughout) planets that are close to their host stars cannot form \textit{in~situ} since the mass of the disk interior to $R=1$~AU is too low (Fig.~5). \citet{M+L2016} have also shown, however, that in fully turbulent disks super-Earths can (in principle at least) form farther out, followed by inward migration. 
\begin{figure}[t]
	\centering
	\includegraphics[width=4in]{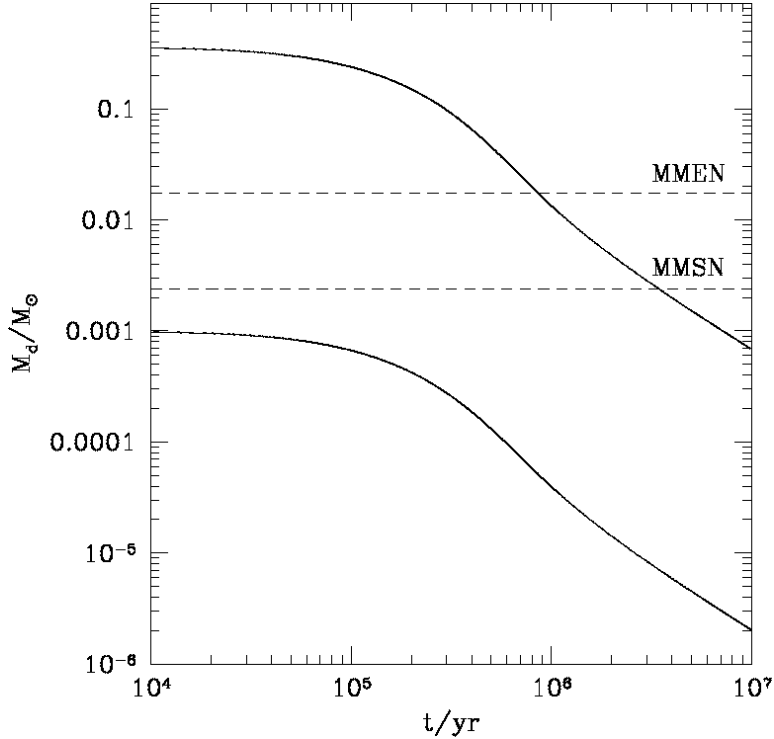}
	\caption{Disk mass up to a radius $R=1$~AU as a function of time (lower curve) and the total disk mass up to 40~AU (upper curve) for a fully turbulent disk model. The infall accretion rate was assumed to decrease exponentially, from an initial $\dot{M}\simeq10^{-5}~M_\odot$ yr$^{-1}$, on a timescale of $t_\textit{ff}=10^{5}$~yr. The dashed lines show the mass at $R<1$~AU for the MMSN and MMEN.\label{f5}}
\end{figure}

Still, it is generally believed that protoplanetary disks are not fully turbulent. Rather, they most likely contain a unionized region of low or no turbulence, known as a ``dead zone'' (e.g., \citep{G+M1998,Simonetal2011}). The dead zone blocks the accretion and allows material to accumulate until it becomes self-gravitating. (Note though that it has been suggested that the Hall effect can ``revive'' the dead zone under certain conditions, e.g., \citep{Lesuretal2014}).

\begin{figure}
	\centering
	\includegraphics[width=4in]{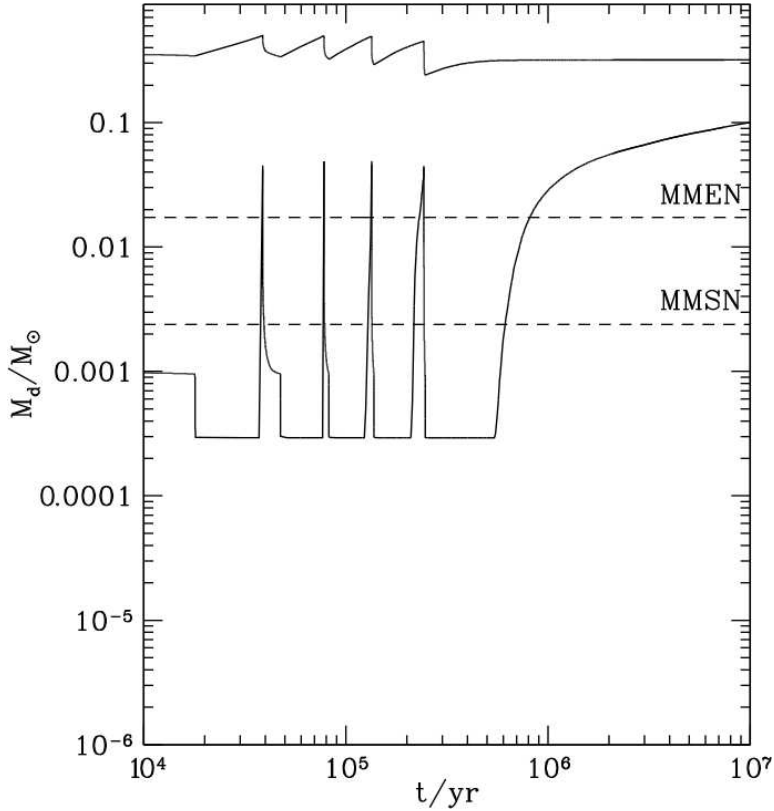}
	\caption{Disk mass up to a radius of $R=1~$AU as a function of time (lower curve) and total disk mass up to a radius of 40~AU (uppercurve). The disk has a dead zone defined by $Re_\mathrm{M,crit}=5\times10^4$. The dashed lines show the mass inside $R=1$~AU for the MMSN (lower) and MMEN (upper).\label{f6}}
\end{figure}

\citet{M+L2016} considered two prescriptions for determining the surface density in the dead zone. In the first, they assumed that the disk surface layers are ionized by external sources to a maximum surface density depth of $\Sigma_\mathrm{crit}/2$ (see discussion in Section~\ref{sec41}). In the second, they assumed that the surface density is determined via a critical magnetic Reynolds number (e.g., \citep{Hawleyetal1995, Flemingetal2000}), $Re_\mathrm{M,crit}$, such that the zone is ``dead'' if $Re_M <Re_\mathrm{M,crit}$. By following the time-dependent evolution of the disk, \citet{M+L2016} showed that at early times the disk undergoes FU~Orionis-type outbursts (see also \citep{Armitageetal2001,Zhuetal2010}). These outbursts occur as the extra heating by self-gravity triggers the magneto-rotational instability within the dead zone (when the infall accretion rate is still high). At later times, as the accretion rate decreases, there are no further outbursts (Fig.~\ref{f6}). Planets that survive must form after the cessation of outbursts, otherwise they are likely to be swept into the star during accretion episodes. Through their time-dependent numerical simulations, \citet{M+L2016} have demonstrated that depending on the dead zone parameters, the disk surface density and the mass inside of 1~AU can build up to several times that of the MMEN (eq.~5), and therefore that the formation of super-Earths in this region is possible (Fig.~\ref{f6}).

\subsection{Application to the Solar System\label{sec41}}
	
There are two possible explanations for the fact that the solar system does not contain any super-Earths: (1)~Either the conditions in the solar nebula were such that super-Earths could not have formed, or (2)~Super-Earths did form but were later removed by some mechanism. In the latter possibility, it is highly unlikely that the removal mechanism was that of ejection through planet-planet scattering, since the average eccentricity in the solar system is low, and the planets are quite coplanar. Consequently, if super-Earths had indeed formed in the solar system, they must have been ``swallowed'' by the Sun.

There is another constraint that determines the likely formation site. As I noted earlier, if the super-Earths had formed outside the snow line, unless they migrated on a timescale shorter than 0.01--0.1~Myr, they would have shepherded the rocky material interior to their orbit, thereby depleting the Earth's formation zone. This would have made the Earth (and other terrestrial planets) volatile---rich, and more similar to a water world than to the current Earth. Consequently, if they formed at all, super-Earths in the solar system likely formed in the innermost regions, inside Mercury's orbit.

This suggests a way (in principle) to solve both the lack of super-Earths problem and the lack of close-in planets in the solar system through a single mechanism: super-Earths could form close to the Sun, clear the inner region of debris, and then fall into the Sun following migration through the gas disk. For this chain of events to actually happen requires some degree of \textit{fine-tuning}. Specifically, the surface density in the active (turbulent) layer must be sufficiently large for the planets to migrate into the Sun, but at the same time small enough to allow the planets to form \textit{in~situ} to begin with. For a super-Earth to migrate into the Sun, it must do so on a timescale shorter than the time it takes the disk to accrete (the viscous timescale). By equating the timescale for type~I migration (obtained when the planet is not massive enough to open a gap in the disk) with the viscous timescale in the disk, \citet{M+L2016} found that the minimum surface density for the planet to migrate into the Sun is given by 
\begin{equation}
\Sigma_\mathrm{min}=940.5\left( \frac{\alpha}{0.01} \right)
\left( \frac{H/R}{0.05} \right)^4
\left( \frac{M}{M_{\odot}} \right)^2
\left( \frac{R}{5~\mathrm{AU}} \right)^{-3/2}
\left( \frac{a}{1~\mathrm{AU}} \right)^{-1/2}
\left( \frac{M_p}{5~M_{\oplus}} \right)^{-1}~\mathrm{g~cm}^{-2}~~.
\end{equation}
Here $\alpha$ is the viscosity parameter, $H/R$ is the disk aspect ratio, $M$ is the mass of the star, $R$ is the radial distance from the central star, $a$ is the orbital radius of the planet and $M_p$ is the planet's mass.

Generally, it is expected that the surface layers of the disk are ionized by cosmic rays or x-rays from the central star (e.g., \citep{M+P2003,Glassgoldetal2004}), to a maximum surface density depth of $\Sigma_\mathrm{crit}/2$ (on each of the disk surfaces). Therefore, if $\Sigma_\mathrm{crit}>\Sigma_\mathrm{min}$, super-Earths formed \textit{in~situ} will migrate into the Sun at the end of the disk's lifetime. If, on the other hand, $\Sigma_\mathrm{crit}<\Sigma_\mathrm{min}$, then migration may not be able to allow the super-Earth to be accreted (even if some Type~I migration takes place). Given the sensitive dependence of $\Sigma_\mathrm{min}$ on the disk's aspect ratio, and therefore on its temperature (see eq.~6), \citet{M+L2016} speculated that in the solar system, super-Earths formed close to the inner boundary of the dead zone of a relatively cool disk (small $H/R$). In that case, there was sufficient time for the super-Earths to migrate and be accreted by the Sun. This would explain both the clearing of the region inside Mercury's orbit and the lack of super-Earths. In this scenario, Mercury and Mars would have formed from a relatively narrow annulus of rocky debris in the orbital range of 0.7--1~AU (see also \citep{Hansen2009}), where the annulus is being truncated at its inner edge by the clearing process. This would explain the relatively small masses of these planets.

I should note that an alternative mechanism, known as the ``grand tack,''has been suggested for pushing the super-Earths into the Sun \citep{Walshetal2011,B+L2015}. In this scenario, Jupiter migrates inwards to 1.5~AU before it gets locked into a resonance with Saturn, and then both Jupiter and Saturn move outwards to their current positions. The migration of Jupiter causes the innermost super-Earths to be shepherded into the Sun.

To conclude this section, the lack of super-Earths in the solar system appears to be somewhat puzzling. However, the second characteristic that makes the solar system somewhat special---the fact that there are no planets or debris inside of Mercury's orbit---may not be a coincidence. Super-Earths that formed \textit{in~situ} in the inner region of a disk that contains a dead zone, could have cleared it of all solid material, with the super-Earths subsequently spiraling into the Sun. \citet{M+L2016} showed that for the dead zone to last a sufficiently long time for the super-Earths to form, the surface density in the active (turbulent) layer must satisfy $\Sigma_\mathrm{crit}\lesssim100$~g~cm$^{-2}$. At the same time, for the super-Earths to eventually migrate into the Sun during the final accretion process, $\Sigma_\mathrm{crit}$ needs to be sufficiently large, so that there would still be sufficient material in the disk. To satisfy both of these constraints (and for the dead zone to last throughout the entire disk lifetime), requires the disk to be sufficiently cool during late accretion. The necessary level of fine-tuning is not excessive, but it still makes the solar system somewhat special in this respect.

An additional element in the solar system that may have played a role in the emergence and evolution of life on Earth is the existence of an asteroid belt, with its associated characteristics. In the next section I discuss asteroid belts and their possible functions in determining the habitability of terrestrial planets.

\section{The Potential Significance of Asteroid Belts for Life\label{sec5}}

There are several ways in which the presence of an asteroid belt in a solar system can (in principle at least) affect the habitability of a terrestrial planet:
\begin{itemize}
	\item[(i)]Since terrestrial planets tend to form in the dry regions of protoplanetary disks~\citep{M+L2012}, water must be later delivered to the rocky surface, and one of the possible mechanisms is through asteroid impacts (e.g., \citep{Morbidellietal2000}).
	\item[(ii)]Large moons can stabilize the rotation axis of planets against chaotic motion, thereby preventing weather extremes. The formation of such moons may again require asteroid or planet impacts (e.g., \citep{C+A2001,Rufuetal2017}).
	\item[(iii)]Life itself or its ingredients may have been delivered to Earth by asteroids (e.g.,  \citep{E+C2000,Houtkooper2011}).
	\item[(iv)]Since the early Earth was molten, gravitational settling resulted in the Earth's crust being depleted of heavy elements such as iron and gold. Some of those are essential for life and they were probably brought to the crust by asteroids (e.g., \citep{Willboldetal2011}).
	\item[(v)]Finally, on a more speculative note, the dominance of mammals and the emergence of intelligent life on Earth might not have happened were it not for the asteroid impact that brought about the extinction of the dinosaurs~\citep{Alvarezetal1980}.
\end{itemize}

While all of these potential effects are somewhat uncertain, even if only one of them is operative, it makes the study of the formation and evolution of asteroid belts important for understanding life on a terrestrial planet. 

The asteroid belt in our own solar system is composed of millions of irregularly shaped bodies made of rock, ices, and metals. It is located between the inner terrestrial planets and the outer giant planets, and its total current mass is about $5\times10^{-4}~M_\oplus$, with about 80\% of the mass being contained in the three largest asteroids, Ceres, Pallas, and Vesta.

Observations and models of the solar system suggest that at the time of planetesimal formation, the snow line---the radial location outside which ice forms---was located inside the asteroid belt~\citep{Abeetal2000,Morbidellietal2000}. In particular, while the asteroids in the inner part of the belt are dry, those more distant than about 2.7~AU (from the Sun) are icy \mbox{C-class} objects. It is generally believed that the asteroid belt is the result of gravitational perturbations caused by Jupiter. Those perturbations did not allow planetesimals to merge and grow, resulting in violent collisions producing fragmentation rather than fusion (e.g., \citep{E+A2004}). 

Since giant planets likely form outside the snow line~\citep{Moralesetal2011}, because the density of solid material there is much higher (due to water ice condensation~\citep{Pollacketal1996}), \citet{M+L2013} proposed that \textit{asteroid belts} (if they form at all) \textit{should be located around the snow line}. To test their hypothesis, \citet{M+L2013} calculated the expected location of the snow line in protoplanetary disk models and compared their results with observations of warm dust in exoplanetary systems, since those may indicate the location of exo-asteroid belts.

\subsection{Water Snow Lines and Asteroid Belts\label{sec51}}

The water snow line marks the distance from the star exterior to which ice forms. It is thought to occur at a temperature of about 170K (e.g., \citep{Lecaretal2006}). In an extended region down to about 100K, the snow border, icy and dry planetesimals can coexist~\citep{M+C2011}.

Particles that migrate through the disk accumulate near the snow line, in a region of relatively small radial extent. This allows them to grow through collisions. Consequently, the formation rate of planetesimals increases by an order of magnitude or more when crossing the snow line (since the solid surface density doubles).

\begin{figure}
	\centering
	\includegraphics[width=4.5in]{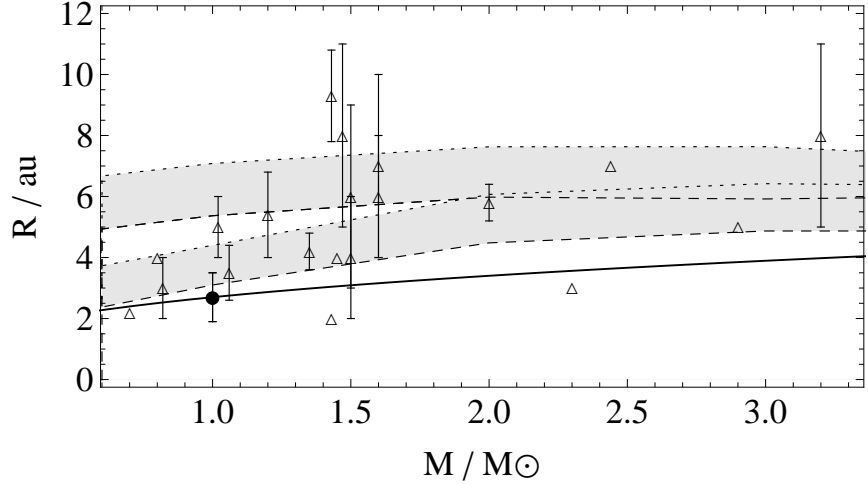}
	\caption{Radius of the observed warm dust. The shaded regions show the snow border found from numerical models at times $t=10^6$~yr (upper) and $t=10^7$~yr (lower). The solid line shows the analytic approximation to the water snow line, given by equation~(7). The open triangles show the individual systems in Table~1. Filled circle shows the location of our solar system's water snow line and the range indicates the extent of our asteroid belt.\label{f7}}
\end{figure}

\citet{M+L2013} modeled the evolution of the water snow line in a layered protoplanetary disk with a dead zone (e.g., \citep{Armitageetal2001,Zhuetal2010}). They found that the snow line moves inward over time, but that its location is only weakly dependent on the mass of the central star. Figure~\ref{f7} shows the radius $R$ of the inner and outer edges of the water snow border found from numerical simulations (corresponding to temperatures of $T=170$K and 100K, at times $t=10^6$~yr and $10^7$~yr). The functional dependence on the stellar mass can be heuristically derived as follows: The temperature that is obtained in the disk as a result of accretion obeys $T^4\propto M/R^3$. If we scale this to the radial location of the water snow line in the solar system we obtain:
\begin{equation}
R_\mathrm{snow}\simeq 2.7\left( \frac{M}{M_{\odot}}\right) ^{1/3}~\mathrm{AU}~~.
\end{equation}

\begin{figure}
	\centering
	\includegraphics[width=4.5in]{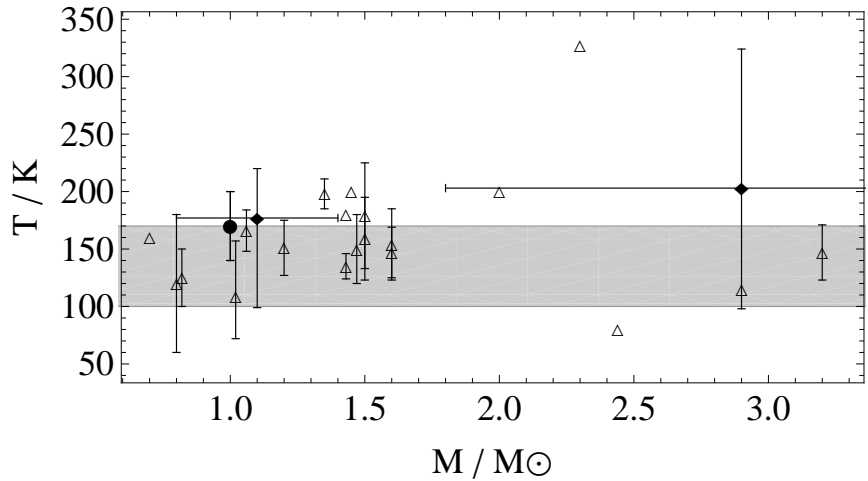}
	\caption{The temperature of the observed warm dust. The shaded region marks the snow border. Open triangles show the individual systems in Table~1. Filled diamonds show two samples from \citet{Moralesetal2011}. Filled circle indicates our solar system.\label{f8}}
\end{figure}

To compare the theoretical predictions with observations, \citet{M+L2013} used observations of debris disks that have a warm infrared component, that could be attributed to an asteroid belt. Table~1 lists a compilation of such debris disks with their determined temperature and inferred radius. Also included in the table are the median temperature values for two samples from \citet{Moralesetal2011}, for which radii were not determined. The radii and temperatures of these putative asteroid belts are also shown in Figures~\ref{f7} and~\ref{f8}, respectively. As we can see from the figures, both the radii and the temperatures agree with the numerical models of the snow line. This supports the proposed scenario of \citet{M+L2013}, in which the location of asteroid belts is around the water snow line.

\begin{table}[t]
	\centering
\caption{Observations of warm dust belts that may be exo-asteroid belts.}
\scriptsize{
\begin{tabular}{lccccccc}
\\
\hline
\noalign{\smallskip}
Source ID &Name &Spectral type &$M$ &$R_\mathrm{dust}$ &$T_\mathrm{dust}$ &Age &Reference\\
&&&$(M_\odot)$ &(au) &(K) &(Myr)\\
\noalign{\smallskip}
\hline
\noalign{\smallskip}
HD 12039 &&G3/5V &l.02 &4--6 &109 &30 &1\\
HD 13246 &&F8V &1.06 &3.5$\pm$0.9 &166$\pm$18 &30 &2\\
HD 15115 &&F2 &1.5\phantom{0} &4$\pm$2 &179$\pm$46 &12 &3\\
HD 15745 &&FO &\llap{(}1.6) &6$\pm$2 &147$\pm$22 &12 &3\\
HD 16743 &&FO/F2III/IV &\llap{(}3.2) &8$\pm$3 &147$\pm$24 &10--50 &3\\
HD 22049 &$\epsilon$ Eri &K2V &0.82 &3$\pm$1 &100--150 &\llap{8}50 &4\\
HD 30447 &&F3V &\llap{(}1.5) &6$\pm$3 &159$\pm$36 &30 &3\\
HD 38678 &$\zeta$ Lep &A2 IV-V(n) &2.3\phantom{0} &3 &327 &\llap{2}31 &5\\
HD 53143 &&G9V/K1V &0.8\phantom{0} &4 &120$\pm$60 &\phantom{0}1 &6\\
HD 53842 &&F5V &1.20 &5.4$\pm$1.4 &151$\pm$24 &30 &2\\
HD 86087 &HR 3927 &AOV &2.44 &7 &\phantom{0}80 &50 &6\\
HD 98800 &&K4/5V &\llap{(}0.7) &2.2 &160 &10 &7\\
HD 109085 &$\eta$ Corvi &F2V &1.43 &2 &180 &\llap{10}00 &6\\
HD 113766 &&F3/F5V &(1.5/1.4) &4 &200 &16 &6\\
HD 152598 &&FOV &l.43 &9.3$\pm$1.5 &135$\pm$11 &210$\pm$70 &2\\
HD 169666 &&F5 &1.35 &4.2$\pm$0.6 &198$\pm$13 &\llap{21}00 &2\\
HD 172555 &HR 7012 &A51V-V &2.0\phantom{0} &5.8$\pm$0.6 &200 &12 &6\\
HD 181296 &$\eta$ Tel &AOVn &2.9\phantom{0} &5 &115 &12 &6\\
HD 192758 &&FOV &\llap{(}1.6) &7$\pm$3 &154$\pm$31 &40 &3\\
HD 218396 &HR 8799 &A5V &1.5\phantom{0} &8$\pm$3 &150$\pm$30 &30--160 &8, 9\\
\noalign{\smallskip}
\multicolumn{5}{l}{Samples from \citet{Moralesetal2011}, median values (range)}\\
\noalign{\smallskip}
\multicolumn{2}{l}{19 solar-type stars} &GOV (KOV-F5) &(l.l (0.8-1.4)) &&177 (99-220) &270 (40-900)\\
50 A-type stars &&AOV (B8-A7) &(2.9 (1.8-3.8)) &&203 (98-324) &100 (5-1000)\\
\noalign{\smallskip}
\hline
\noalign{\medskip}
\end{tabular}
\textit{Note.} The masses in parentheses have been derived from the spectral type. References. (1)~\citet{Hinesetal2006}; (2)~\citet{Mooretal2009}; (3)~\citet{Mooretal2011}; (4)~\citet{Backmanetal2009}; (5)~\citet{Moerchenetal2010}; (6)~\citet{Chenetal2006}; (7)~\citet{Lowetal2005}; (8)~\citet{Chenetal2009}; (9)~\citet{MoroMartinetal2010}. References 1--9 in the table correspond (sequentially) to references [57]--[65] in the reference list.
}
\end{table}

\subsection{Giant Planet Location\label{sec52}}

\begin{figure}
	\centering
	\includegraphics[]{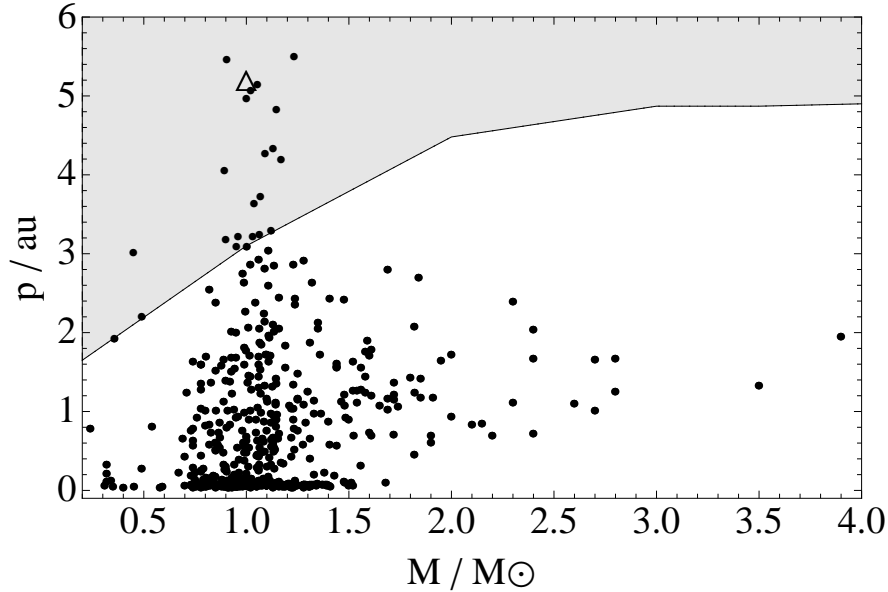}
	\caption{The distribution of observed giant planet periastron separation, $p$, against the mass of the central star, $M$. The open triangle shows where Jupiter lies. The shaded region shows the icy domain outside of the lower limit to the water snow line predicted by numerical models.\label{f9}}
\end{figure}

Figure~\ref{f9} shows the periastron separation for 520 giant planets (with masses larger than $10~M_\oplus$) as a function of the central star's mass~\citep{Wrightetal2011}. Also shown in the figure is the lower limit of the distances of the water snow line obtained in the numerical models of \citet{M+L2013}. The region exterior to the water snow line is shaded. The planets that are observed to be close to their host star are thought to have migrated inwards through a gas-rich disk~\citep{L+P1986,Haischetal2001}. Only giant planets that form when the gas is already considerably depleted can avoid migration. Simulations show that the conditions required for a gas giant around a Sun-like star to linger around Jupiter's orbital distance are obtained only in about 1--2 percent of the systems~\citep{Armitageetal2002}.

In systems in which giant planet migration occurs, we do not expect to find substantial or compact asteroid belts, since the asteroids are scattered to larger distances, or are accreted by the planet or the central star~\citep{F+N2007}. The observed warm dust belts listed in Table~1, therefore, likely remained intact because the giant planets in these systems migrated little or not at all. For this non-migration to have happened, the giant planets must have formed towards the end of the lifetime of the protoplanetary disk.

Figure~\ref{f9} shows that only 19 out of 520 giant planets (less than 4\% of these observed systems) are located outside the water snow line. These statistics suggest that only a small fraction of the observed systems contain a compact asteroid belt, making our solar system somewhat special. I should note, however, that the observed statistics are almost certainly affected by selection effects, since planets with larger orbital separations are more difficult to detect (see e.g., \citep{Cummingetal2008} for a discussion).

To conclude this part, in our solar system Jupiter may have migrated only by about 0.2--0.3~AU~\citep{Morbidellietal2010}. The asteroid belt was probably much more massive initially, consisting of about one Earth mass. Due to Jupiter's migration, however slight, most of this mass has been ejected, leaving behind only about 0.001 of the original mass. This course of events may have also been important for the emergence and evolution of life on Earth, since had the asteroid belt remained very massive, the number of impacts on Earth (due to the continued perturbations from Jupiter) might have been too high to allow for the evolutionary processes to follow their course. For example, any planet around Tau Ceti (which may be orbited by five planets) would experience many more impacts than Earth, due to a much more massive debris disk in that system. It may therefore be that the time interval during which the giant planet should form is rather restricted, if it is to allow complex life to emerge and evolve. Too much migration may altogether disrupt the asteroid belt, and too little may produce far too many devastating impacts. This conclusion assumes, of course, that some asteroid impacts are indeed necessary for life. Interestingly, asteroids are presently regarded as a potential threat for humanity in the future, and NASA has been testing a computer program called \textit{Scout} that is supposed to act as a celestial intruder alert system, to warn against incoming asteroids. In that sense, asteroids can giveth and asteroids can taketh away.

Given that not only the individual impacts themselves, but also their \textit{rate} can have significant effects on the habitability of terrestrial planets, it is important to examine a few of the elements that determine this rate. Specifically, a high rate of asteroid impacts can render a highly cratered planet, not hospitable for life. On the other hand, too low a rate could suppress the delivery of elements that are essential for the emergence and evolution of life.

\subsection{Asteroid Impacts on Terrestrial Planets\label{sec53}}

\subsubsection{Introduction\label{sec531}}

As I noted earlier, if asteroid belts form at all, they most likely form around the location of the water snow line, interior to giant planets. Throughout the time that the gas disk still exists, the eccentricities and inclinations of the asteroids are damped by tidal interactions with the protoplanetary disks (e.g., \citep{Ward1989,Artymowicz1993,K+I2002}). The lifetime of the gas disk is typically a few million years~\citep{Haischetal2001}, after which it is being dispersed by photoevaporation (see discussion in \citep{Armitage2013}). After the gas disk is removed, gravitational perturbations clear asteroids at many resonance locations (e.g.,  \citep{Gladmanetal1997,M+M2010,Chrenkoetal2015}). This clearing creates potential Earth (and other terrestrial planets) impactors. In the solar system Jupiter and Saturn, which are the largest planets, are the main drivers of the dynamical evolution of the asteroid belt; even though the effects that Jupiter had on the collision rate with the Earth is debated~\citep{H+J2008,H+J2012}. Similarly, giant planets can be the driving force of evolution (if an asteroid belt exists) in exoplanetary systems. 

Resonances that play a major role in the asteroid belt dynamics are \textit{mean-motion} resonances and \textit{secular} resonances. In a mean-motion resonance, the ratio of the orbital periods of two objects is a ratio of two integers. Several of the known mean-motion resonances in the solar system, such as the Kirkwood gaps \citep{D+M1983,Moons1996,OBrienetal2007}, are found within the asteroid belt. Secular resonances arise when the apsidal precession rate of two objects orbiting a common central object are close to each other (e.g., \citep{F+S1986, Yoshikawa1987}). From our perspective here, the most important secular resonance in the solar system is the $\nu_6$ resonance (\citep{Bottkeetal2002,M+M2011} and references therein). It relates the apsidal precessions of the asteroids and Saturn and it sets the inner boundary (at about 2~AU) of the solar system's asteroid belt. Each resonance has a certain libration width in semi-major axis, over which it is effective. Asteroids that fall within a libration width undergo perturbations that cause their eccentricities to increase to the point where they are either ejected from the system, or they collide with a planet or the central star. Regions in which libration widths overlap are dynamically chaotic regions~\citep{M+H1999}, and almost all asteroids are cleared from such regions. The outer edge of the asteroid belt in the solar system (at about 3.3~AU) is determined by the overlapping of Jupiter's resonances.

\begin{figure}[t]
	\centering
	\includegraphics[width=\textwidth]{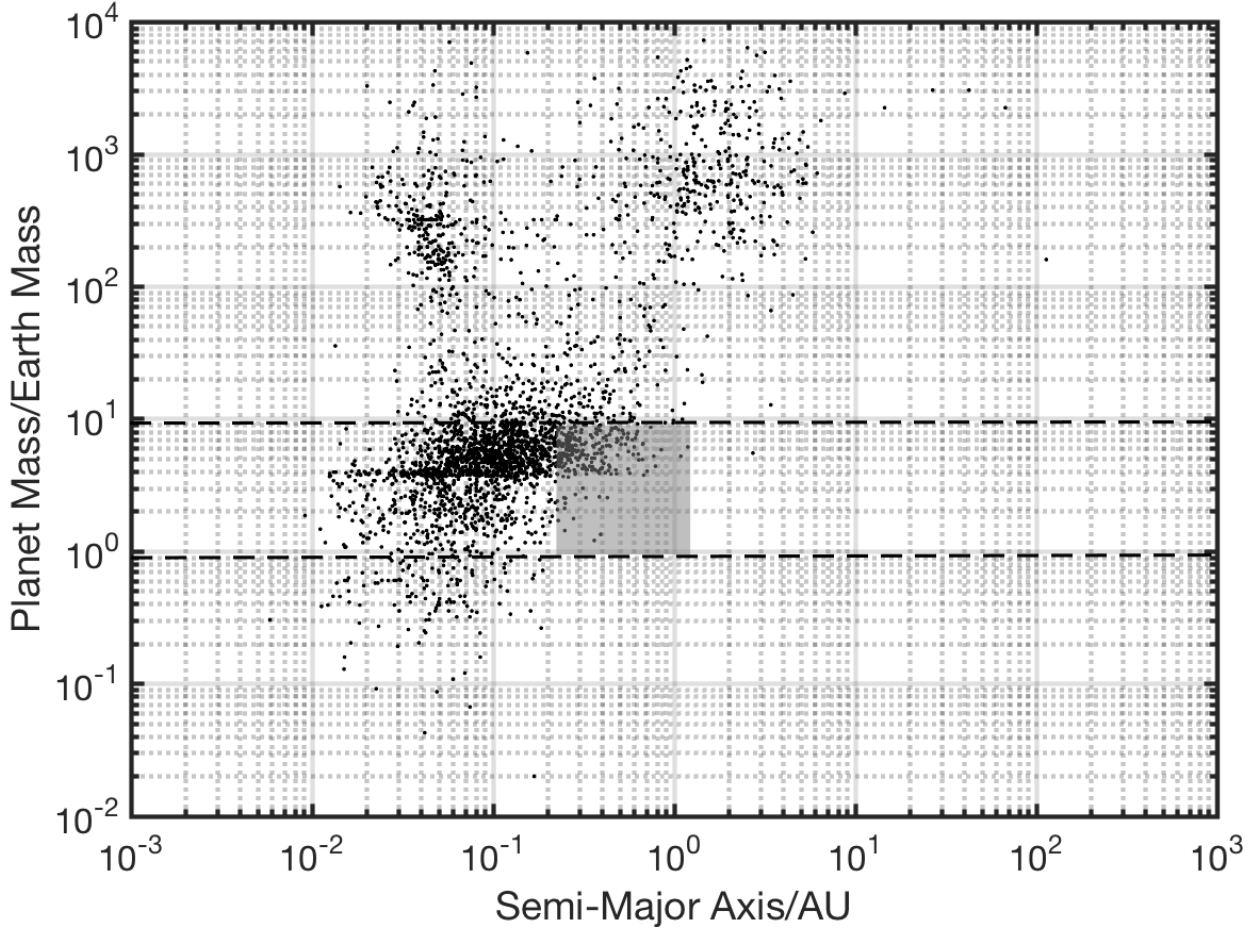}
	\caption{Planet masses and semi-major axes of observed exoplanets (from \citep{Hanetal2014}). The area between the dashed lines contains the range of super-Earth masses used in the simulations described in Sections~\ref{sec532}--\ref{sec534}. The transparent gray box highlights the observed super-Earths with a semi-major axis corresponding to the inner solar system.\label{f10}}	
\end{figure}

The fact that many of the exoplanetary systems contain a super-Earth (e.g., \citep{JontofHutteretal2015}; Fig.~\ref{f10}) raises the additional question of how the presence of a super-Earth in the solar system might have affected the rate of asteroid impacts on Earth. Smallwood et~al.~\citep{Smallwoodetal2018} studied in particular multi-planet systems in which there had been no process (e.g., migration) that could have destroyed the asteroid belt or could have prevented the formation of terrestrial planets. More specifically, they investigated how the architecture of systems such as the solar system affects the rate of asteroid impacts on Earth (or a similar terrestrial planet).

\subsubsection{The Numerical Method: \boldmath{$N$}-Body Simulations\label{sec532}}

\citet{Smallwoodetal2018} used the hybrid symplectic integrator in the orbital dynamics package MERCURY to model the structure of the asteroid belt and the rate of impacts on Earth (see \citep{Chambers1999} for details of the package). The motions of Jupiter, Saturn, a super-Earth, the Earth, and a distribution of asteroids all orbiting a central object were simulated. The asteroids were assumed to gravitationally interact with the star and the planets, but not with each other. The evolution of the orbit of each asteroid was followed for ten million years.

In the solar system, there are more than 10,000 asteroids with high-accuracy measurements of their semi-major axes. The mean of those values is $<\!a\!>=2.74\pm0.616$~AU. The mean eccentricity is $<\!e\!>=0.148\pm0.086$, and the mean inclination $<\!i\!>=8.58^{\rm o}\pm6.62^{\rm o}$ (see description in \citep{M+D2000}). At the same time, the precise initial distribution of the asteroids within the belt (immediately following the dispersal of the protoplanetary gas disk) is not accurately known. In their simulations, \citet{Smallwoodetal2018} therefore assumed a uniform distribution (see also \citep{L+F1997}), with the semi-major axis of each asteroid given by
\begin{equation}
a_i=\left( a_\mathrm{max} - a_\mathrm{min}\right) \times\xi_r + a_\mathrm{min}~~~,
\end{equation}
where $a_\mathrm{min}=1.558$~AU is the inner boundary of the distribution, $a_\mathrm{max}=4.138$~AU is the outer boundary, and $\xi_r$ is a randomly generated number between 0 and 1. The inner and outer boundaries were based on the structure of the solar system, with $a_\mathrm{min}$ being three Hill radii (the region in which a body's gravity dominates) beyond the semi-major axis of Mars, and $a_\mathrm{max}$ being three Hill radii inside Jupiter's orbit. The Hill radius is given by 
\begin{equation}
R_H=a_p\left( \frac{M_p}{3~M_s} \right)^{1/3}~~,
\end{equation}
where $a_p$ is the planet's semi-major axis, $M_p$ is its mass, and $M_s$ is the mass of the star. No asteroids are likely to be located within 3~Hill radii, since this is the planet's gravitational reach (e.g., \citep{Gladman1993,Chatterjeeetal2008,M+M2015}).

The orbit of each asteroid is defined by six orbital elements: (i)~the semi-major axis, $a$, which was taken to be distributed uniformly in the range $a_\mathrm{min} <a <a_\mathrm{max}$; (ii)~the inclination angle $i$, randomly chosen from the range 0--10$^{\rm o}$; (iii)~the eccentricity, $e$, randomly generated from the range 0.0--0.1; (iv)~the longitude of the ascending node, $n$; (v)~the argument of perihelion, (the angle from the ascending node to the object's periastion), $g$; (vi)~the mean anomaly (angular distance from pariastion), $M_a$. The last three elements were uniformly randomly sampled from the range 0--360$^{\rm o}$. Given that the solar system is stable over long timescales~\citep{D+L1998,I+T2002}, the current orbital parameters of the planets were also taken as initial parameters.

To test their numerical scheme, \citet{Smallwoodetal2018} first checked the scalability of the results with respect to the number of asteroids used and with respect to the radius used for the Earth. Based on these tests, they decided to run their simulations with $10^4$ asteroids and with an inflated Earth, with a radius of $2\times10^6$~km (otherwise the number of impacts during the simulation is too low to allow for statistically significant conclusions). Neglecting asteroid--asteroid interactions was fully justified by the fact that the timescale for asteroid--asteroid collisions is much longer (in fact, of the order of the age of the solar system) than the timescale for action by resonance effects (of the order of 1~Myr; e.g., \cite{Dohnanyi1969}).

\subsubsection{The Effects of the Architecture of the Inner Solar System\label{sec533}}

Since \citet{M+L2015} identified the absence of super-Earths as perhaps the most significant characteristic that distinguishes the solar system from other exoplanetary systems, \citet{Smallwoodetal2018} first varied the mass and semi-major axis of an artificially added super-Earth in the inner solar system. The super-Earth was taken to be initially in a circular orbit with zero inclination. The super-Earth's mass was varied in the different simulations in the range 1--10$~M_\oplus$ and its semi-major axis was taken to be in the range 0.2--1.4~AU. In each one of the simulations, \citet{Smallwoodetal2018} followed the dynamics for 10~Myr, and determined the number of asteroid impacts on Earth (which was inflated in radius as described above), the impacts on Jupiter, on Saturn, on the central star, the number of asteroids ejected from the system (achieving semi-major axes larger than 100~AU), and the number of asteroids remaining within the initial distribution of the asteroid belt. All the runs were then compared to that of a solar system without a super-Earth, in order to evaluate the significance of the absence of a super-Earth in our solar system. Figure~11 gives the total number of collisions with Earth during the period of 10~Myr. Overall the trend is that the addition of a super-Earth interior to the Earth's orbit increases the number of asteroid collisions with the Earth.
\begin{figure}
	\centering
	\includegraphics[width=4.5in]{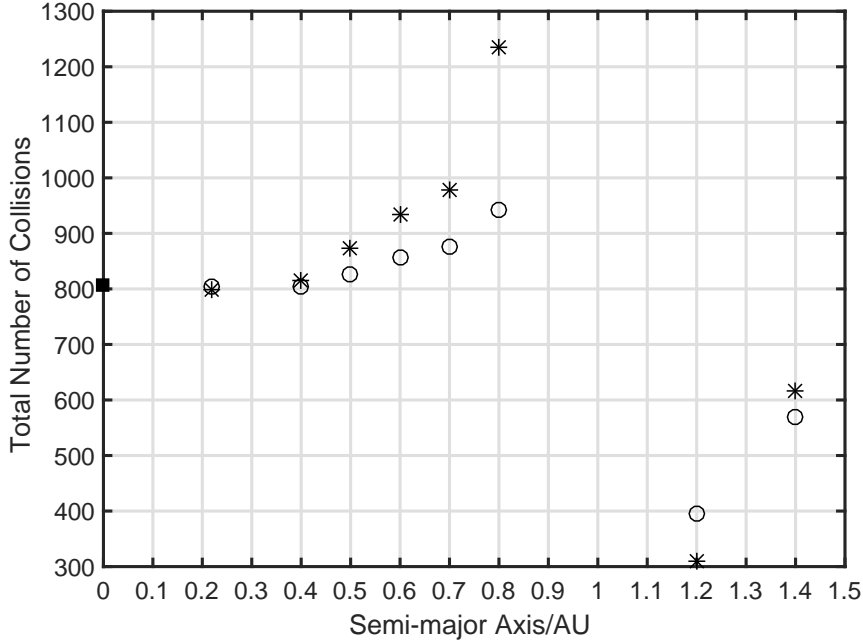}
	\caption{The total number of collisions with the (inflated) Earth as a function of the semi-major axis of the super-Earth. The stars represent simulations with a 10~$M_\oplus$ super-Earth and the circles with a 5~$M_\oplus$ super-Earth. The square represents the simulation without a super-Earth.\label{f11}}
\end{figure}

When the super-Earth is located exterior to the Earth's orbit the total number of impacts onto the Earth is lower than in the absence of a super-Earth. For the parameters used in the simulations, a 10~$M_\oplus$ super-Earth located at a semi-major axis of 0.8~AU produced the largest number of impacts on Earth, whereas a 10~$M_\oplus$ super-Earth located at 1.20~AU caused the lowest number of impacts. At the same time, when the super-Earth was placed at 1.2 and 1.4~AU, the number of asteroids ejected from the system increased significantly. Generally, for \textit{interior} (to the Earth's orbits) super-Earths, the number of impacts was found to increase with increasing distance from the Sun.

\begin{figure}
	\centering
	\includegraphics[width=4.5in]{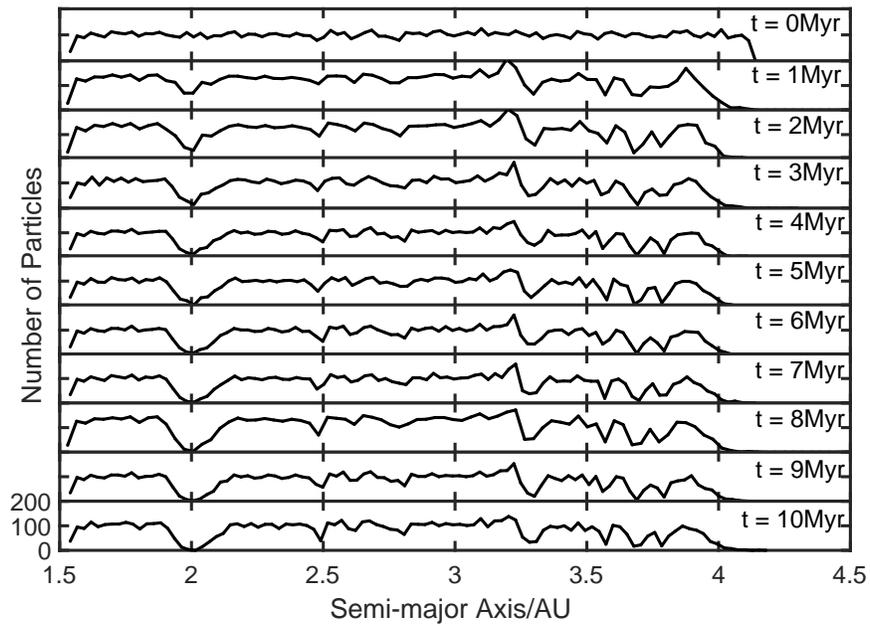}
	\caption{Evolution of the asteroid distribution without a super-Earth.\label{f12}}
\end{figure}
\begin{figure}
	\centering
	\includegraphics[width=4.5in]{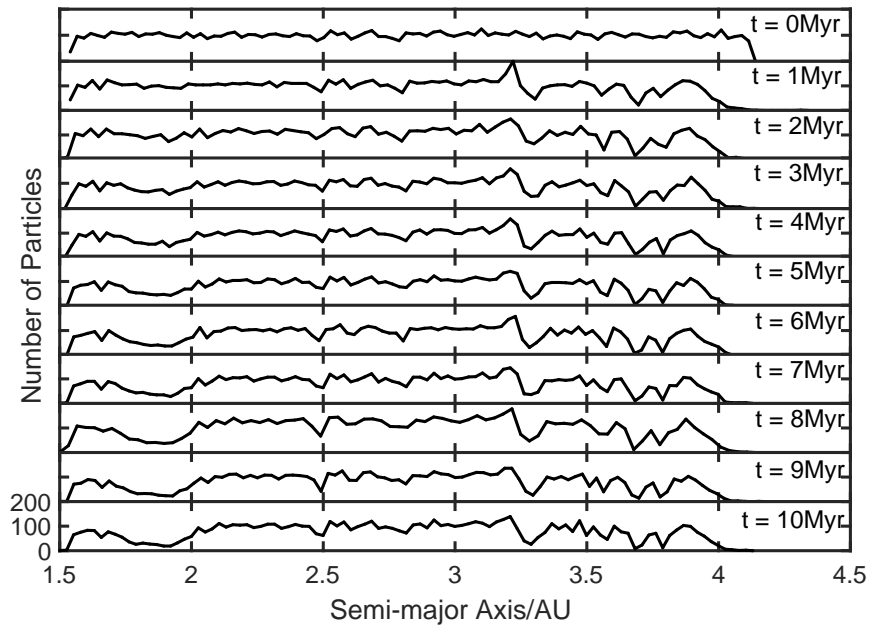}
	\caption{Evolution of the asteroid distribution with a 10~$M_\oplus$ super-Earth located at a semi-major axis of 0.8~AU.\label{f13}}
\end{figure}
\begin{figure}
	\centering
	\includegraphics[width=4.5in]{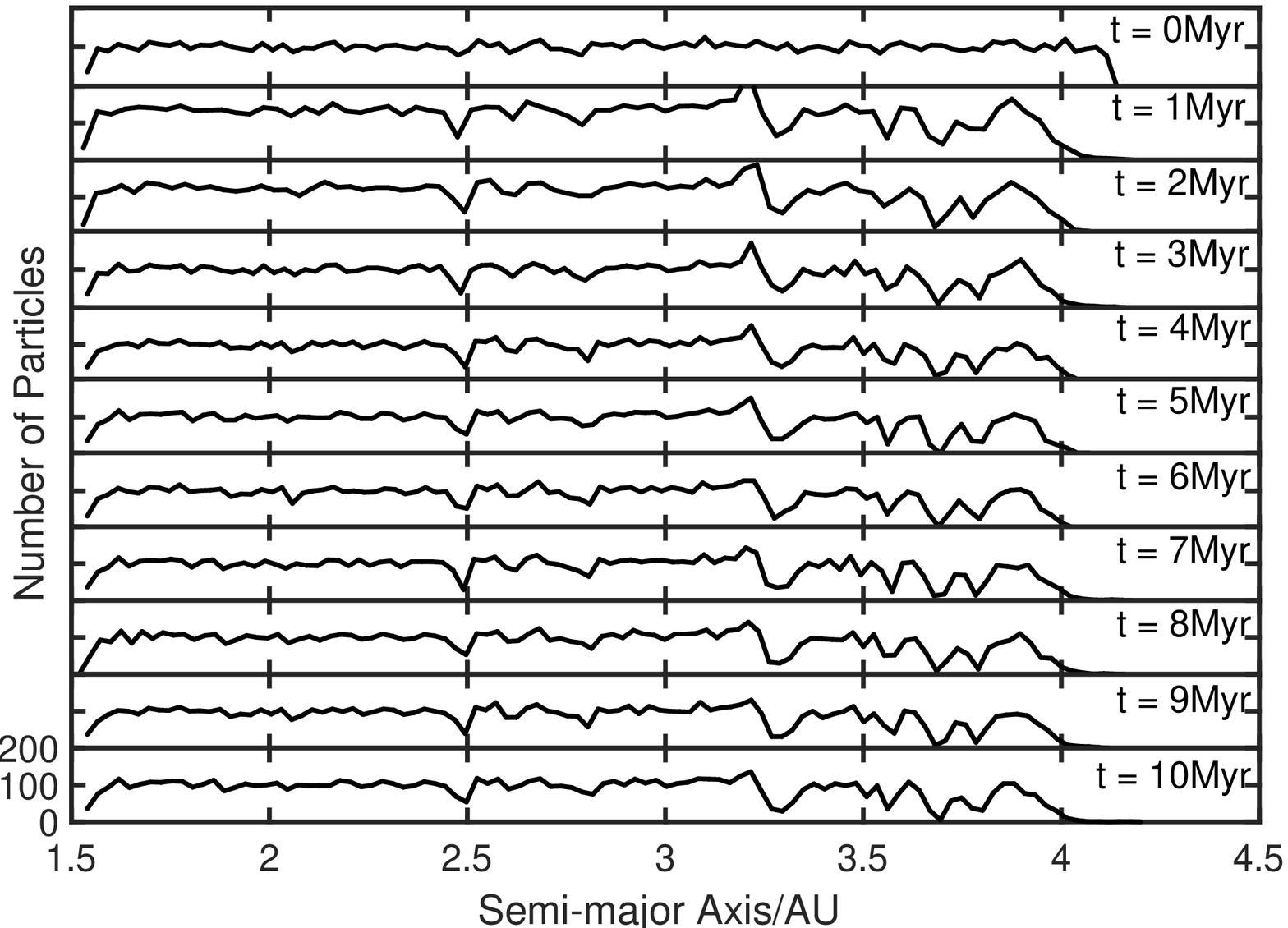}
	\caption{Evolution of the asteroid distribution with a 10~$M_\oplus$ super-Earth located at 1.2~AU.\label{f14}}
\end{figure}

\begin{figure}
	\centering
	\includegraphics[width=4.5in]{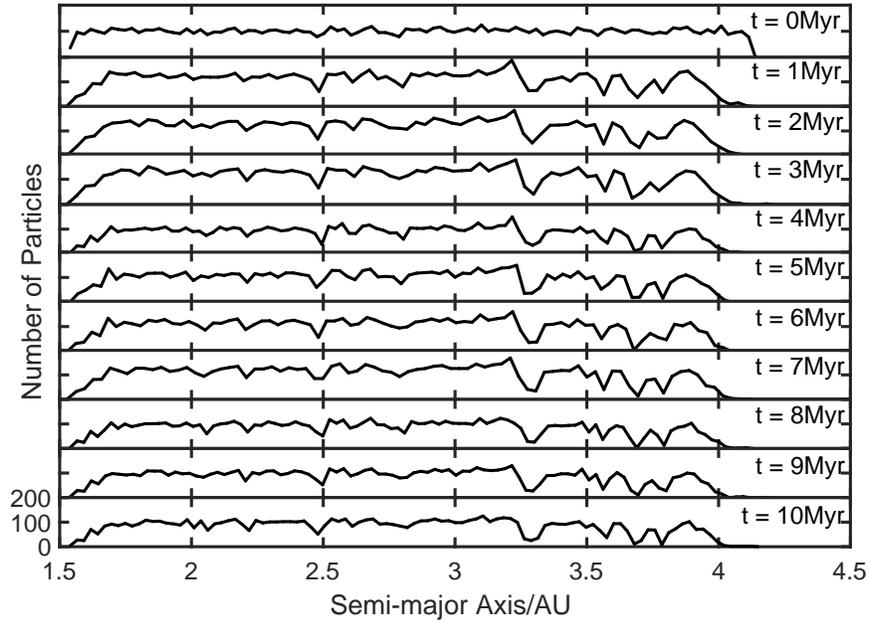}
	\caption{Evolution of the asteroid distribution with a 10~$M_\oplus$ super-Earth at 1.4~AU.\label{f15}}
\end{figure}

	
\citet{Smallwoodetal2018} also examined the evolution of the asteroid belt itself, and this is shown in Figures~\ref{f12}--\ref{f15}. The four panels show a simulation with no super-Earth (Fig.~\ref{f12}), one with a 10~$M_\oplus$ super-Earth at $a=0.8$~AU (Fig.~\ref{f13}), a 10~$M_\oplus$ super-Earth at $a=1.2$~AU (Fig.~\ref{f14}) and a 10~$M_\oplus$ super-Earth at $a=1.4$~AU (Fig.~\ref{f15}). The distribution of the asteroids was calculated every million years for 10~Myr. As time progresses, perturbations caused by mean-motion and secular resonances with Jupiter and Saturn clear out regions in the asteroid distribution. The most important mean-motion resonances are the 3:1, 5:2, 7:3, and 2:1. These are located at 2.5~AU, 2.8~AU, 2.9~AU, and 3.3~AU respectively. Overlapping libration widths of the mean-motion resonances produce Jupiter's chaotic region, which is located from about 3.6~AU to the outer edge of the asteroid distribution (at about 4.13~AU). The resonance that plays the most important role in determining the number of collisions with Earth is the $\nu_6$ resonance, which I will discuss in detail below.
\begin{figure}
	\centering
	\includegraphics[width=4.5in]{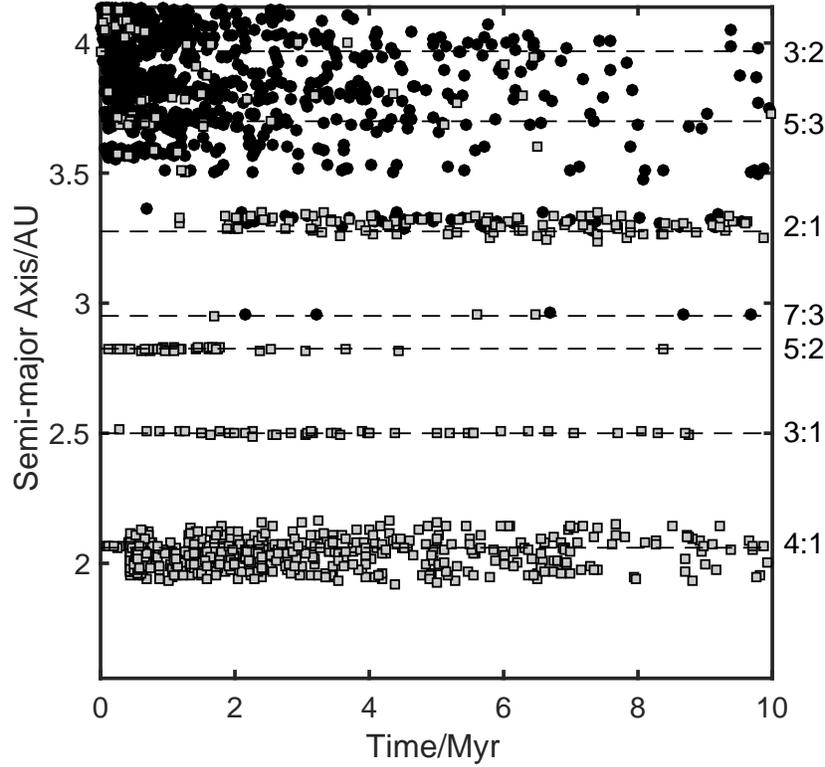}
	\caption{The original semi-major axis of each asteroid as a function of the time when the final outcome occurred, for a system with no super-Earth. The outcomes depicted are collisions with the Earth (gray squares) and ejections (black circles). The mean-motion resonances with Jupiter are represented by the dashed lines.\label{f16}}
\end{figure}
\begin{figure}
	\centering
	\includegraphics[width=4.5in]{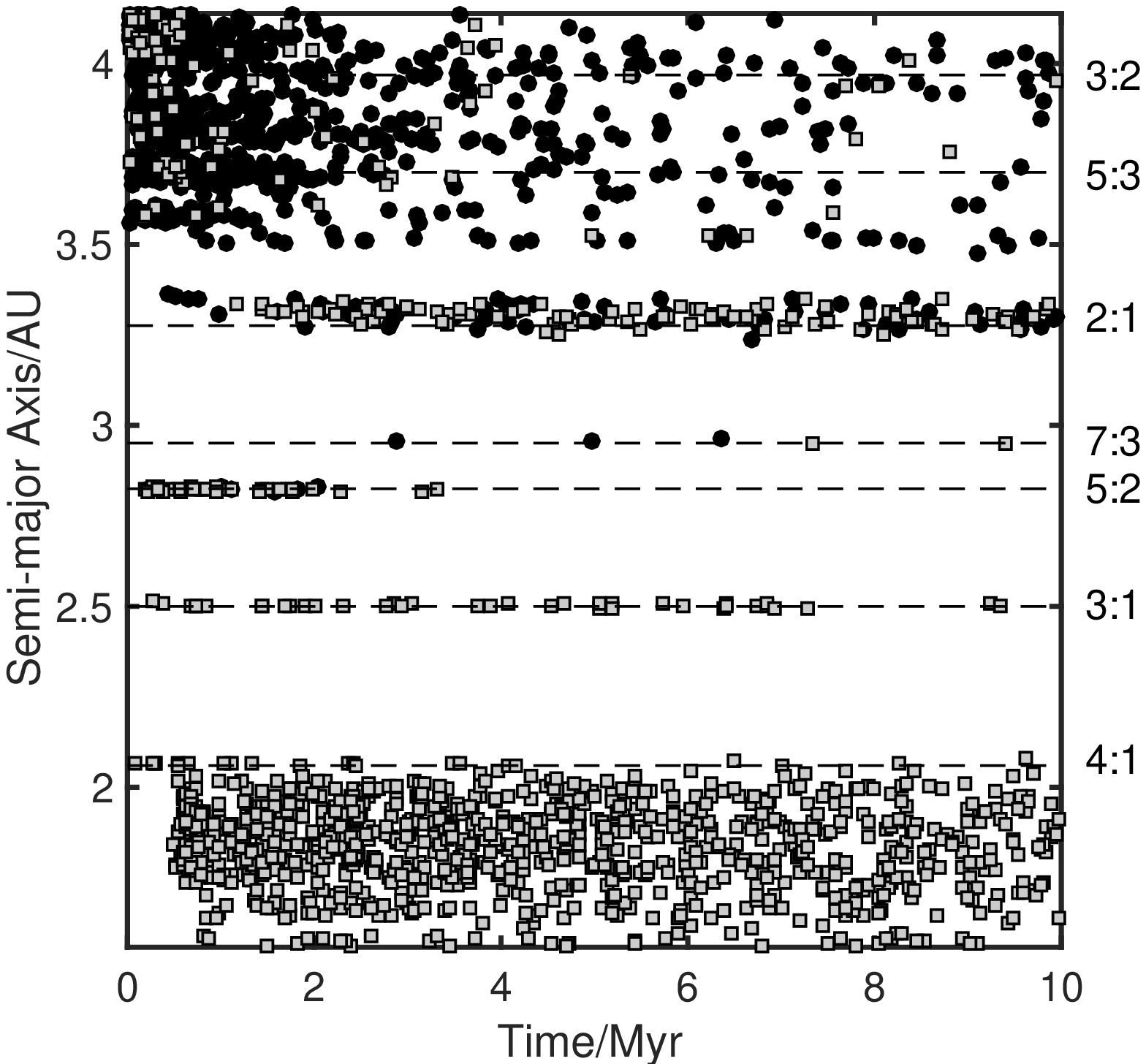}
	\caption{Same as Figure~\ref{f16}, with a super-Earth located at 0.8~AU.\label{f17}}
\end{figure}
\begin{figure}
	\centering
	\includegraphics[width=4.7in]{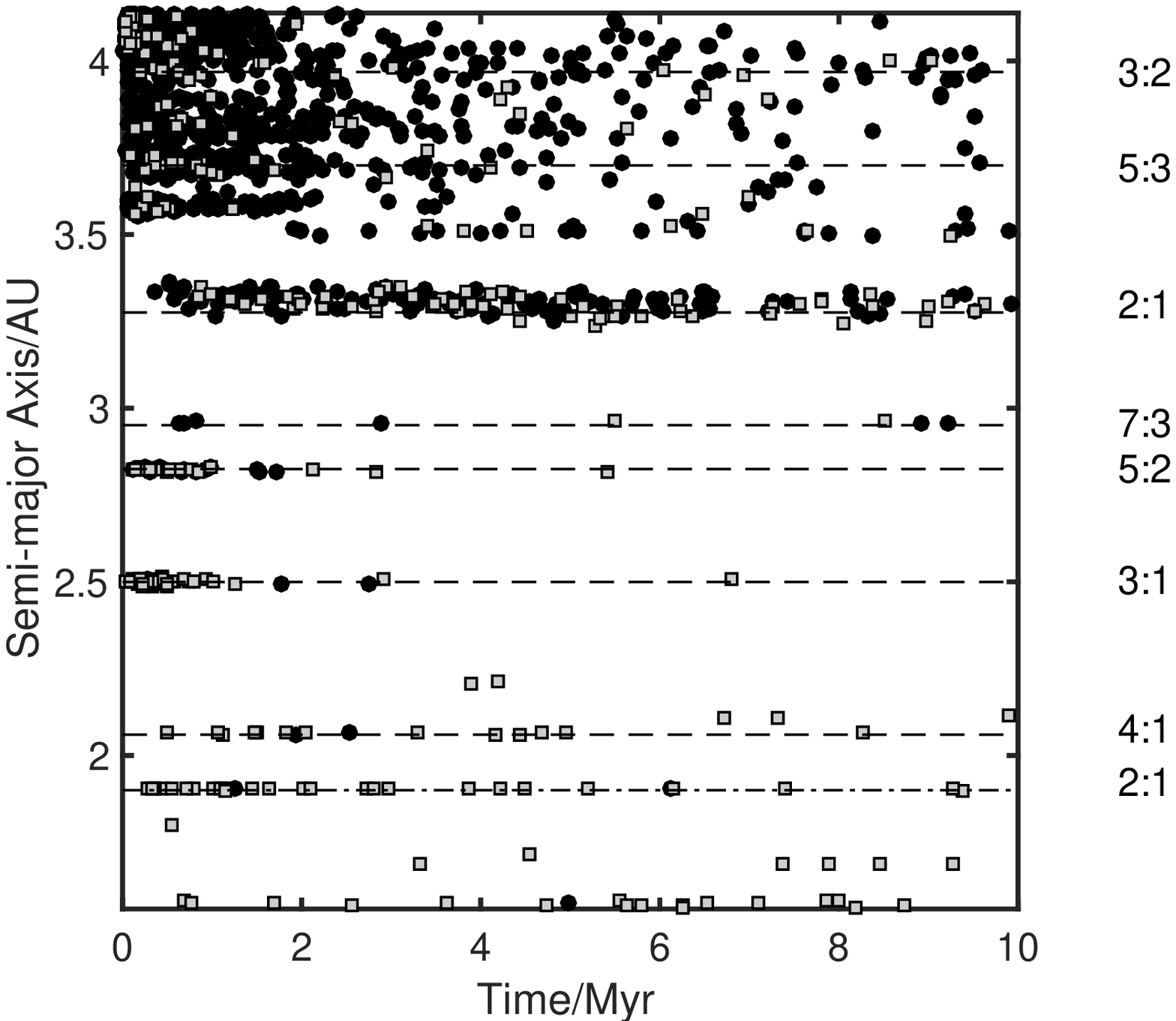}
	\caption{Same as Figure~\ref{f16}, with a super-Earth located at 1.2~AU.\label{f18}}
\end{figure}
\begin{figure}
	\centering
	\includegraphics[width=4.7in]{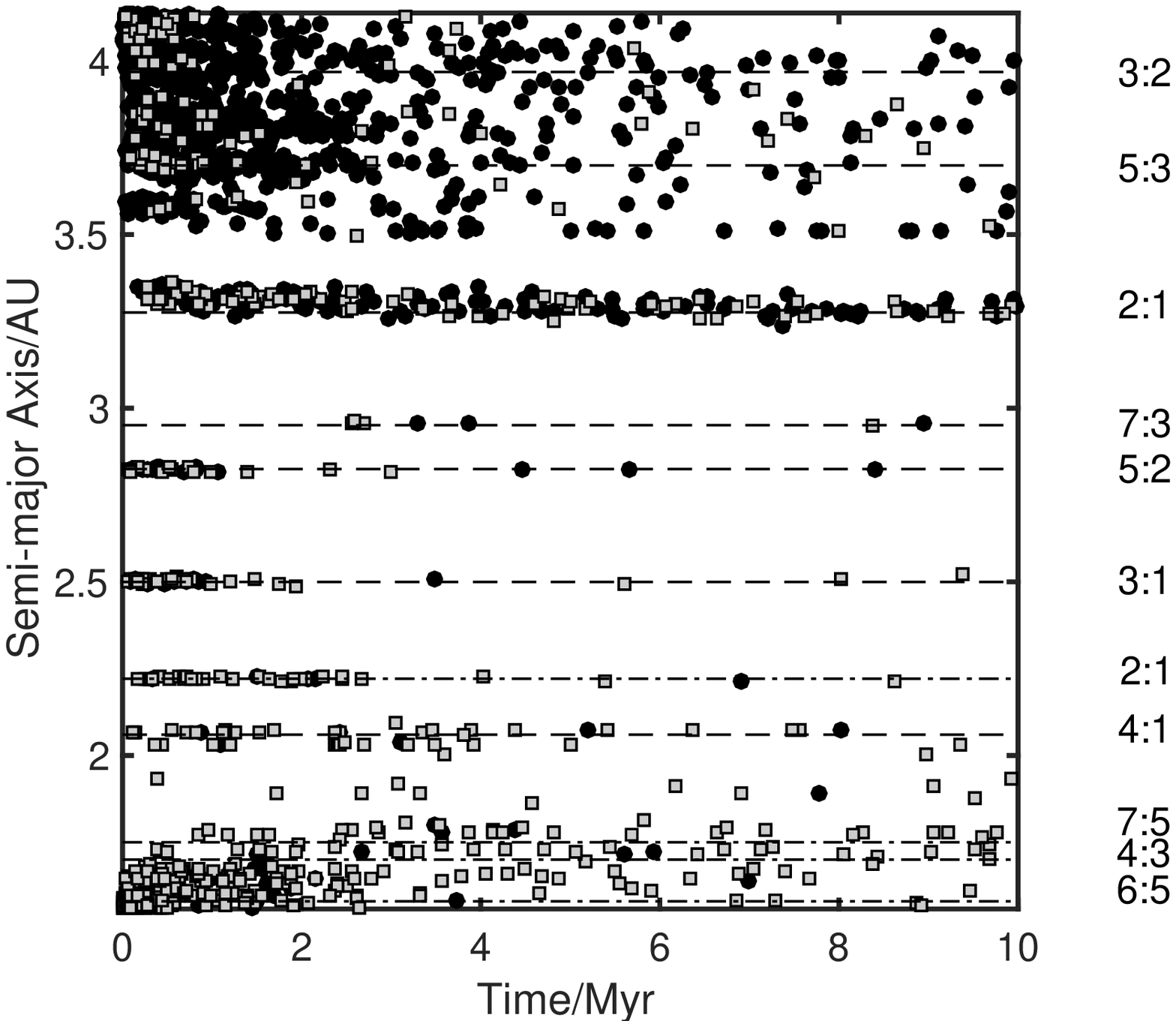}
	\caption{Same as Figure~\ref{f16}, with a super-Earth located at 1.4~AU.\label{f19}}
\end{figure}

There are five potential outcomes for the fate of each asteroid during the various simulations: It can be ejected from the solar system, impact the Earth, collide with another planet, collide with the Sun, or remain in the asteroid belt. The ejections and impacts with the Earth are shown in Figures~\ref{f16}--\ref{f19}, by the black circles and gray squares, respectively. Figure~\ref{f16} shows the results for a system with no super-Earth. Figure~\ref{f17} contains a 10~$M_\oplus$ at a semi-major axis of $a=0.8$~AU. Figure~\ref{f18} has a 10~$M_\oplus$ super-Earth at $a=1.2$~AU and Figure~\ref{f19} a 10~$M_\oplus$ super-Earth at $a=1.4$~AU. The figures show the \textit{initial} semi-major axis for each asteroid (that is, its point of origin) as a function of the time of its final outcome. Asteroids that were originally at resonance locations are cleared out because their eccentricities increase through the action of the mean-motion and secular resonances. The locations of the mean-motion resonances between Jupiter and the super-Earth are shown on the vertical axis at the right-hand side. Placing a 10~$M_\oplus$ super-Earth at $a=0.8$~AU widens the $\nu_6$ secular resonance, thereby increasing the number of asteroids perturbed into Earth-impacting orbits. When the super-Earth is placed exterior to Earth's orbit, two effects act to decrease the number of impacts on Earth: (i)~the $\nu_6$ resonance is suppressed, and (ii)~a 2:1 mean-motion resonance is created within the asteroid belt, which acts to cause additional clearing out of asteroids. When a 10~$M_\oplus$ super-Earth is placed at 1.4~AU, it creates a chaotic zone which is produced by the overlapping of the libration widths of the super-Earth's 6:5, 4:3, and 7:5 mean-motion resonances. In turn, this chaotic zone clears a large number of asteroids from the inner parts of the asteroid belt. The chaotic region can be clearly seen in Figure~\ref{f19}. 

\begin{figure}
	\centering
	\includegraphics[width=.49\textwidth]{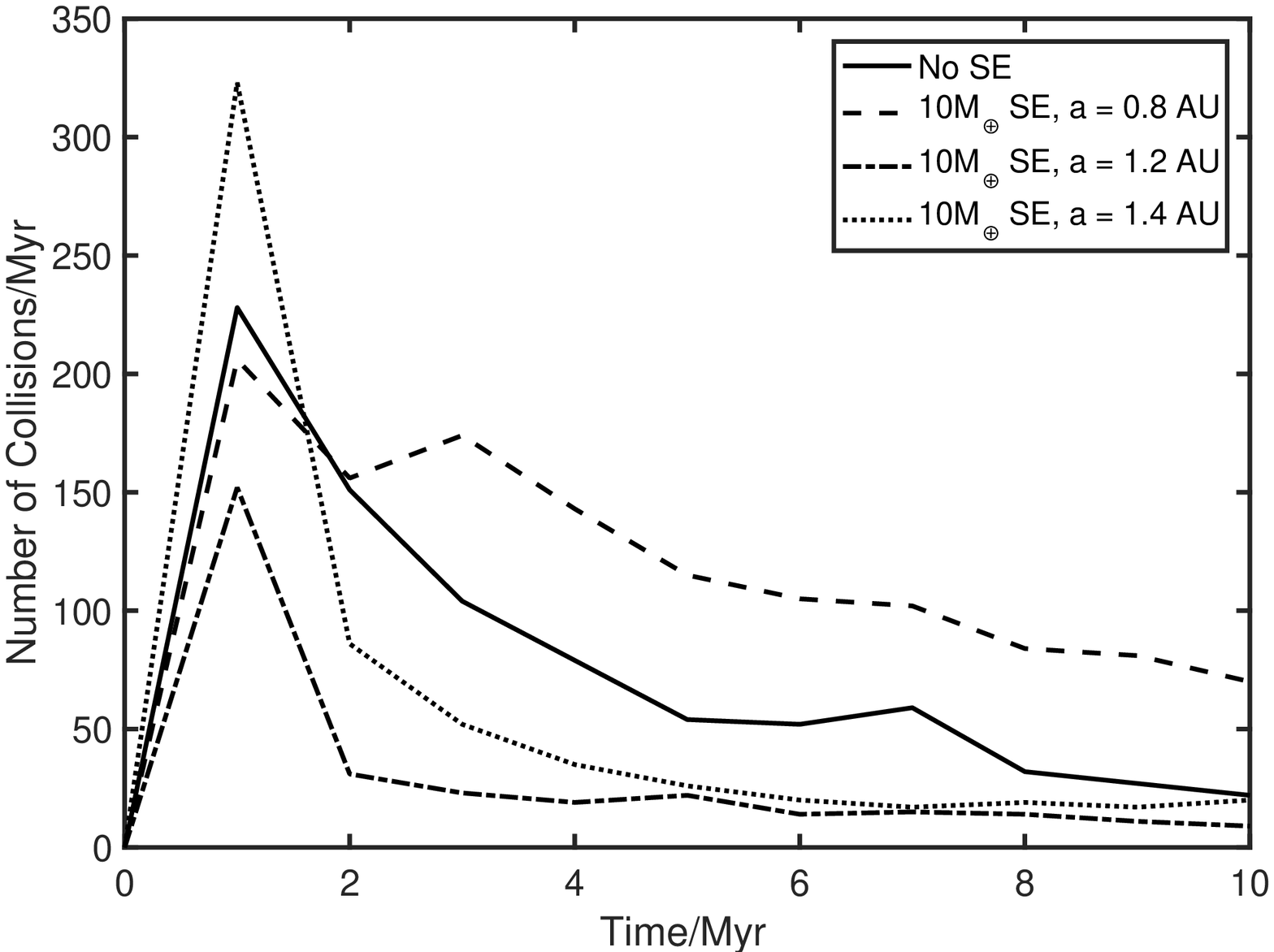}~
	\includegraphics[width=.49\textwidth]{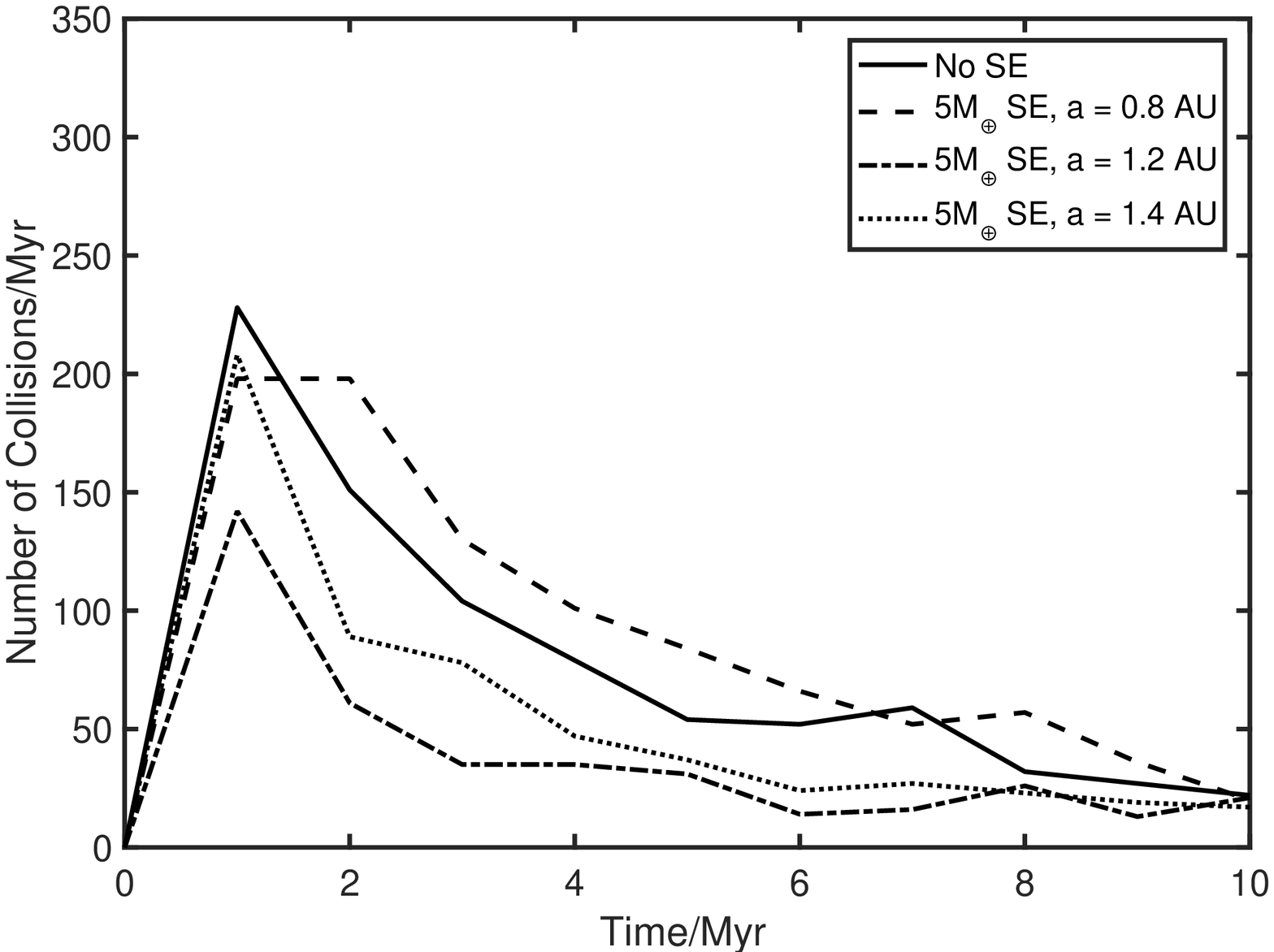}
	\caption{The collision rate with Earth per million years. The left panel involves a 10~$M_\oplus$ super-Earth and the right panel a 5~$M_\oplus$ super-Earth.\label{f20}}
\end{figure}

The effects of the mass of the super-Earth on the rate of asteroid impacts on Earth is shown in Figure~\ref{f20}. What is depicted is the number of impacts per million years for a super-Earth with a mass of 10~$M_\oplus$ (left panel) and a mass of 5~$M_\oplus$ (right panel), for various semi-major axis values. The initial spike in the impact rate is due to the fact that the number of asteroids in the belt is larger at the beginning of the simulations than at their end. The fact that a 10~$M_\oplus$ super-Earth at 1.4~AU produces the highest impact rate at 1~Myr is due to the chaotic region created by the super-Earth. The rate however rapidly declines as asteroids are cleared out. Over the 10~Myr duration covered by the simulation, a super-Earth located at 0.8~AU produces the highest rates, both for a 10~$M_\oplus$ super-Earth and a 5~$M_\oplus$ one.

\begin{figure}
	\centering
	\includegraphics[width=.49\textwidth]{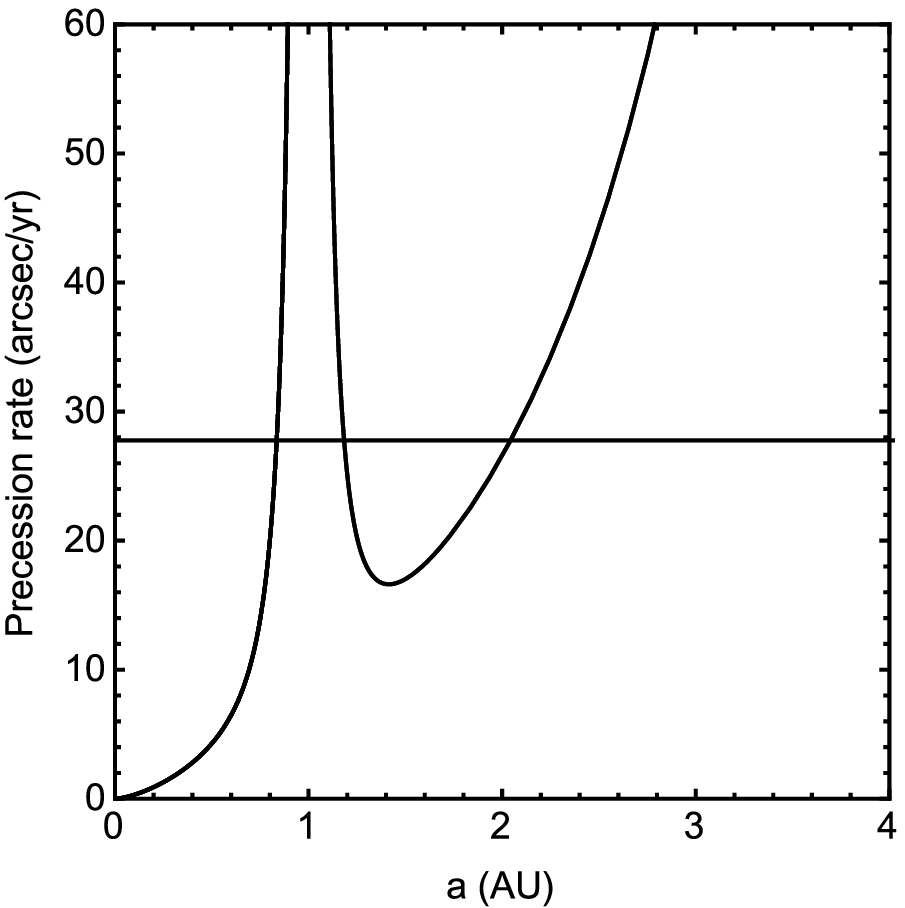}~
	\includegraphics[width=.49\textwidth]{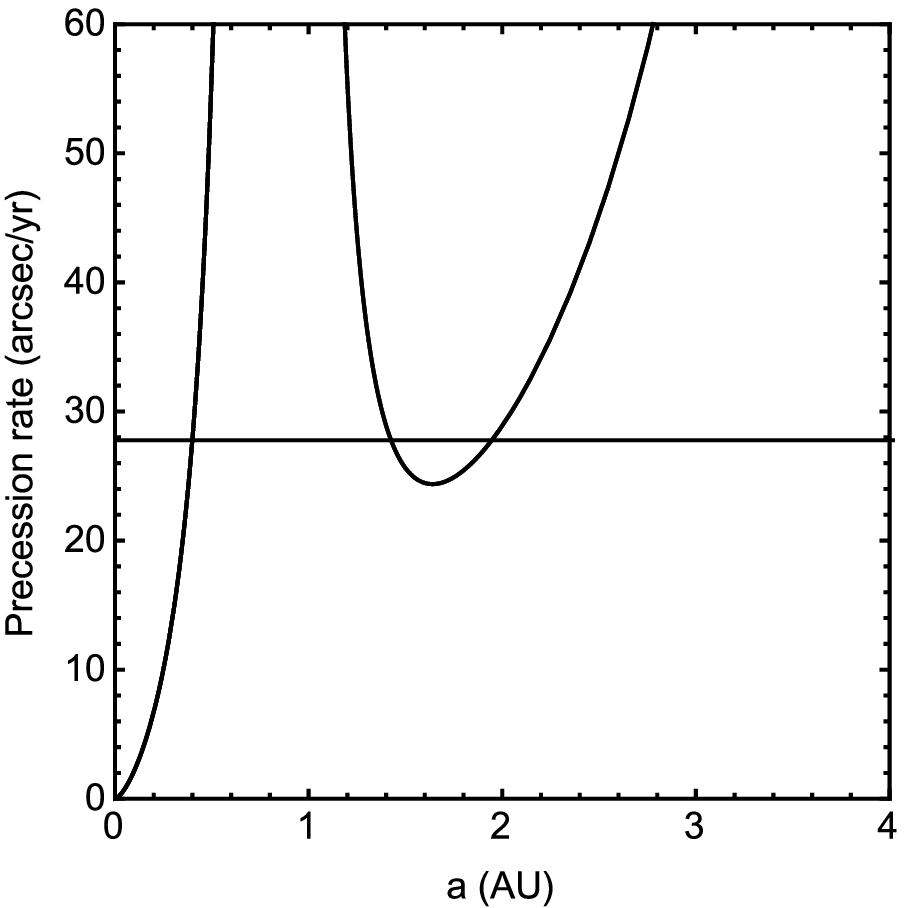}\\
	\includegraphics[width=.49\textwidth]{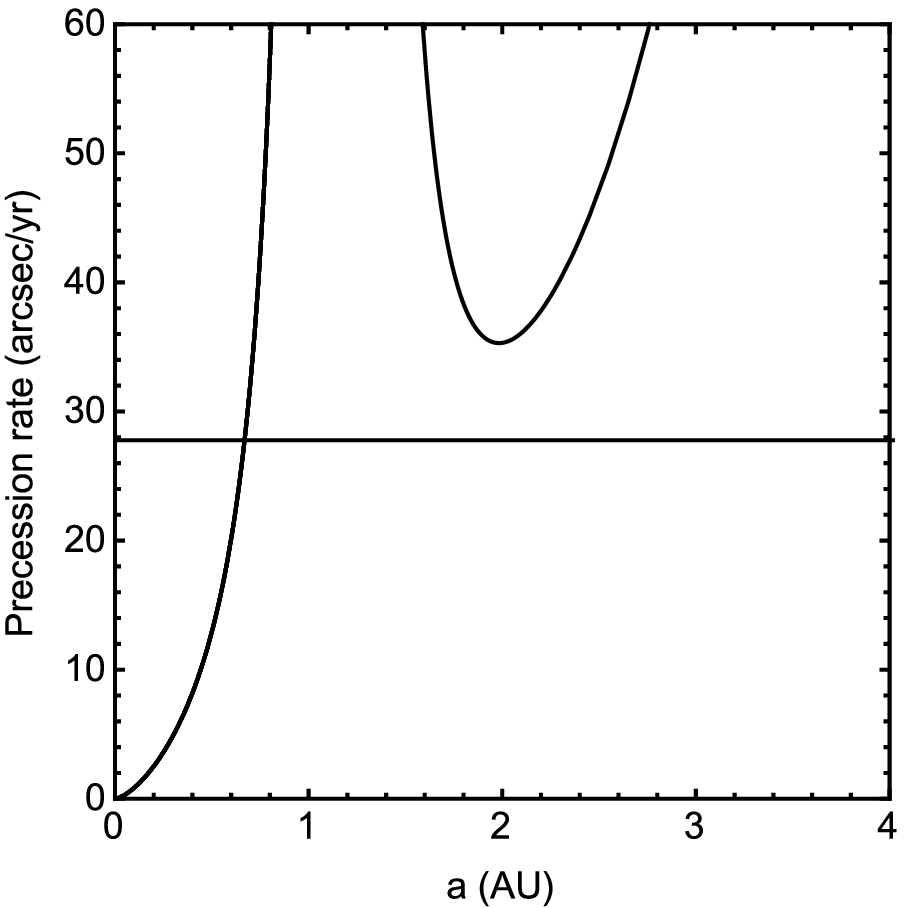}~
	\includegraphics[width=.49\textwidth]{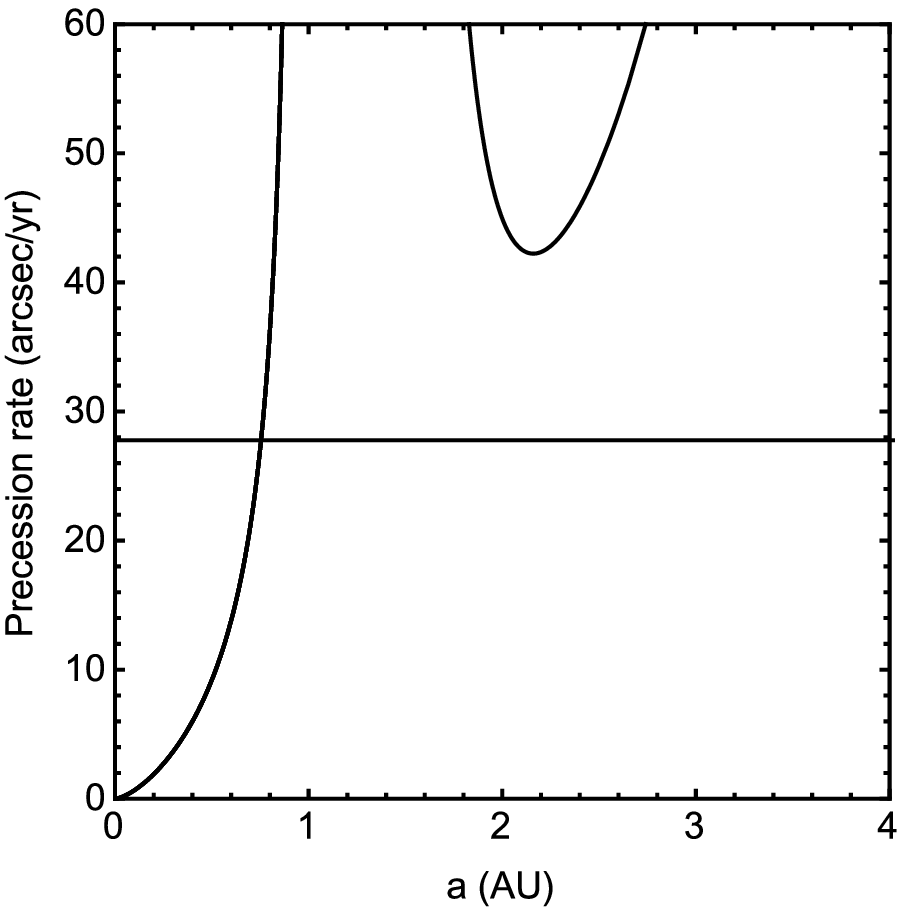}
	\caption{The precession rate of a test particle as a function of semi-major axis. The horizontal line represents the $g_i$ eigenfrequency of Saturn. The intersection of the precession rate with the eigenfrequency denotes the location of a secular resonance. The top-left panel is for a system with no super-Earth. Top-right: a 10~$M_\oplus$ super-Earth at $a=0.8$~AU. Bottom-left: a 10~$M_\oplus$ super-Earth at $a=1.2$~AU. Bottom-right: a 10~$M_\oplus$ super-Earth at $a=1.4$~AU.\label{f21}}
\end{figure}

As I have noted earlier, the $\nu_6$ resonance is a major contributor to the rate of asteroid impacts on Earth. In fact, the majority of the asteroids colliding with the Earth originate from the location of the $\nu_6$ resonance in the asteroid belt. Specifically, in the simulations of \citet{Smallwoodetal2018} without a super-Earth, the total number of asteroid impacts on Earth produced by secular resonances was about 2.5 times higher than that produced by mean-motion resonances. \textit{If} asteroid impacts were indeed important for the emergence and/or evolution of life on Earth, then the $\nu_6$ resonance may have played a significant role in our planet's habitability. The $\nu_6$ resonance involves both Saturn and Jupiter. Basically, Jupiter increases the precession frequency of the asteroids, so that they fall into a resonance with the apsidal precession rate of Saturn. In Figure~\ref{f21} (taken from \citep{Smallwoodetal2018}), I show the precession rate of a test particle as a function of orbital separation. The solid horizontal line marks the eigenfrequency of Saturn (found using a generalized form of secular perturbation theory; e.g., \cite{M+D2000}). The top left panel of Figure~\ref{f21} represents our solar system since it includes Earth, Jupiter, Saturn, and the asteroid belt. The intersection of the particle's precession rate with Saturn's eigenfrequency represents the location of the $\nu_6$ resonance, at about 2~AU. 

The introduction of a super-Earth can change the asteroid precession rate so as to either enhance or altogether remove the resonance with Saturn, depending on the super-Earth's location. For example, when the super-Earth has a semi-major axis of 0.8~AU, the precession rate (of the test particle) is close to Saturn's eigenfrequency for semi-major axis values in the range 1.5--2.0~AU, which enhances the $\nu_6$ resonance (top right panel in Fig.~\ref{f21}). On the other hand, when the super-Earth is exterior to Earth's orbit, the $\nu_6$ resonance is removed---the precession rate of the test particle doesn't intersect Saturn's eigenfrequency (bottom panels; the left is for a super-Earth at 1.2~AU and the right at 1.4~AU). This behavior agrees with that observed in Figures~\ref{f16}--\ref{f19}, where an interior super-Earth produced a widening of the libration width of the $\nu_6$ resonance, while an exterior super-Earth led to the disappearance of the resonance. The agreement of the numerical results with the behavior expected from the generalized form of the secular perturbation theory (represented in Figure~\ref{21}) gives great confidence in the numerical simulations.

\subsubsection{The Effects of the Architecture of the Outer Solar System\label{sec534}}

\begin{figure}
	\centering
	\includegraphics[width=4.5in]{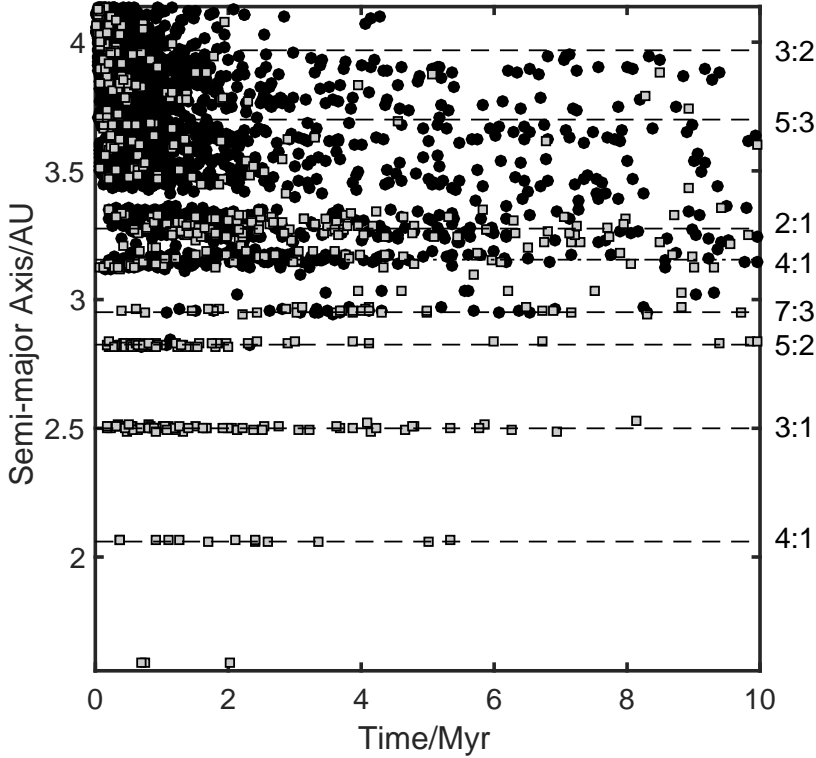}
	\caption{The original semi-major axis of each asteroid as a function of the time when the final outcome occurred, when Saturn is located at $a=8$~AU. The dash-dotted line marked 4:1 represents a mean-motion resonance between the asteroids and Saturn. The dashed lines represent mean-motion resonances with Jupiter. Gray squares denote impacts on Earth and circles ejections.\label{f22}}
\end{figure}

\begin{figure}
	\centering
	\includegraphics[width=4.5in]{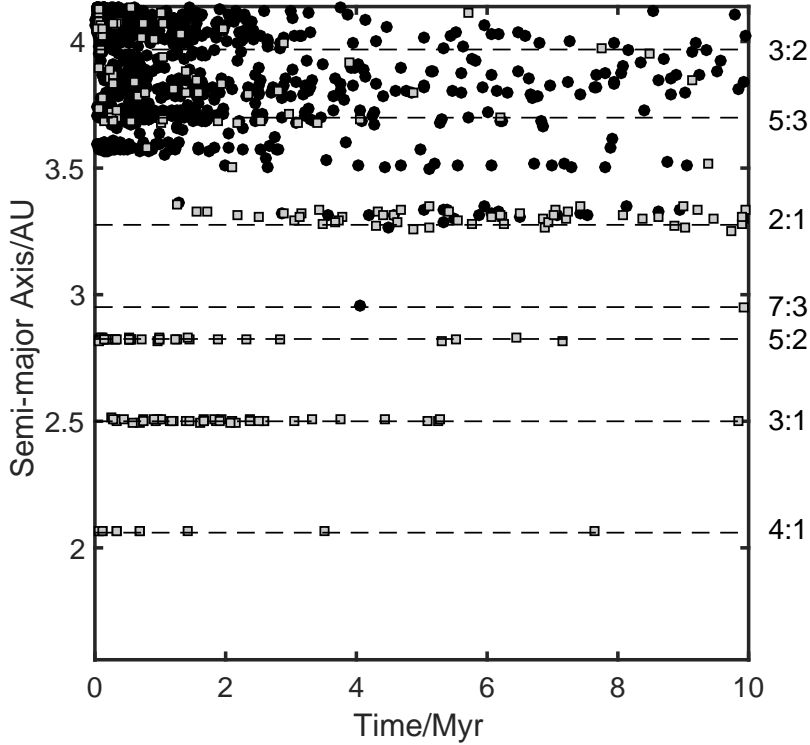}
	\caption{The same as Figure~\ref{f22}, with Saturn at $a=12$~AU.\label{f23}}
\end{figure}

In the next step, \citet{Smallwoodetal2018}  considered the effects of the orbital properties of the giant planets. They found that the location of the $\nu_6$ resonance is rather insensitive to changes in Saturn's mass, moving outward only slightly as the mass is increased. The resonance was found to be much more sensitive to changes in Saturn's orbital separation from the Sun. Specifically, as Saturn is moved outward, the resonance location moves inward. However, I should note that Saturn's orbital location may not be accidental, since it is close to a 5:2 resonance with Jupiter. To further investigate the effects of the architecture of the giant planets, \citet{Smallwoodetal2018} ran additional simulations, while varying Saturn's mass and semi-major axis. They found that increasing Saturn's semi-major axis (from its nominal $\sim$9.5~AU) results in the location of the $\nu_6$ resonance moving eventually outside the boundaries of the asteroid belt. Decreasing the semi-major axis moved the $\nu_6$ resonance toward the middle (and then outer) region of the asteroid belt.These results are also reflected in Figures~\ref{f22}--\ref{f23}. Figure~\ref{f22} shows the initial semi-major axis for each asteroid as a function of the time of its final outcome, when Saturn's semi-major axis is decreased to be $a=8.0$~AU. Figure~\ref{f23} shows the results when Saturn is taken to be at $a=12.0$~AU. For $a=8.0$~AU, the $\nu_6$ resonance shifts to the outer part of the asteroid belt, which decreases the number of impacts on Earth (compared to  $a=9.5$~AU), but increases the number of ejections. When Saturn's semi-major axis is taken to be 12.0~AU, the $\nu_6$ resonance is located outside the asteroid distribution. Overall, Smallwood et~al.'s simulations~\cite{Smallwoodetal2018} showed that Saturn's orbital semi-major axis has a significant effect on the location of the $\nu_6$ secular resonance, and thereby on the rate of asteroid impacts on Earth.

To conclude this part, it appears that the $\nu_6$ resonance plays an important role in producing asteroid impacts on terrestrial planets in the inner part of a planetary system. The architecture of both the inner part and of the outer part affect the location of the $\nu_6$ resonance and thereby the rate of impacts. Super-Earths with masses larger than 5~$M_\oplus$ and interior to the Earth's orbit would have increased the impact rate, while super-Earths exterior to the Earth's orbit would have decreased it. The position of Saturn in the solar system also had a significant effect on the rate of asteroid impacts on Earth. Significantly changing the semi-major axis of Saturn in both directions would have generally resulted in a decrease in the number of asteroid collisions with the Earth. However, since the orbital location of Saturn is close to being in the 5:2 resonance with Jupiter, it may not be accidental.

\subsection{The CO Snow Line\label{sec54}}

\subsubsection{Introduction\label{sec541}}
The most abundant volatiles in a protoplanetary disk are H$_2$O, CO, and CO$_2$. A snow line marks the location in the disk where the mid-plane temperature is sufficiently low, so that a volatile condenses out of the gas phase and becomes solid. The snow line of each volatile has its own radial distance from the central star, with the water snow line being the closest to the star. As I have noted earlier, giant planets are expected to form outside the water snow line, because the density of solids there is higher (e.g., \citep{Pollacketal1996,R+J2013}). In general, the composition of planets and their atmospheres are largely determined by the location of their formation relative to the snow lines~\citep{Obergetal2011}. While the water snow line is found at a temperature of about 170~K (see Section~\ref{sec51}), the CO snow line occurs at T$_\mathrm{CO, snow}=17$~K~\citep{Obergetal2005}. Comets from the solar system's Kuiper belt show different amounts of CO, suggesting that they formed close to the CO snow line~\citep{AHearnetal2012}. The Kuiper belt is thought to have formed in the region extending from about 27~AU (from the Sun) to about 35~AU~\citep{Levisonetal2008}. Since the CO snow line would have been in this region at the time of planetesimal formation, it could mark the transition from the planet-forming zone to the dwarf planets/small icy bodies-forming zone.

While it is very difficult to detect the water snow line in exosolar systems (because of its relative proximity to the host star), the CO snow line presents an easier target because it is farther away. The best observed extrasolar snow line is in the disk around TW~Hya. This star has a mass of 0.8~$M_\odot$ and an age of less than 10~Myr \citep{Hoffetal1998}. \citet{Qietal2013} detected in the star's vicinity the reactive ion N$_2$H$^+$, which is only present when CO is frozen out. These authors determined the radial distance of the CO snow line to be 28--31~AU, very similar to our own solar system.

Since CO ice is needed to form methanol---a building block of more complex organic molecules---we need to understand the evolution of the CO snow line, because that evolution plays a crucial role in the origin of the prebiotic molecules that had led to the emergence of life on Earth~\citep{WWN2010}. Recall that comets had intensely bombarded the young Earth, apparently delivering those ingredients that were necessary for life. I should note that debris disks---the equivalent of the solar system's Kuiper belt---are extremely common in exosolar systems and hence the solar system in not special in that sense. Similarly, systems that have giant planets that are the equivalent of Jupiter, may be expected to have Oort clouds.

\citet{M+L2014} studied the evolution of the CO snow line using two protoplanetary disk configurations: (i)~a fully turbulent disk model and (ii)~the more likely model for a protoplanetary disk---a disk with a dead zone (low turbulence region).

\subsubsection{A Turbulent Disk Model\label{sec542}}
In fully turbulent disks, material is assumed to orbit at radius $R$ with a Keplerian angular velocity $\Omega=\sqrt{GM_*/R^3}$, where $M_*$ is the star's mass (e.g., \citep{Pringle1981}). The effective viscosity in a disk in which turbulence is driven by the magneto-rotational instability (MRI) is usually parametrized in the form~\citep{S+S1973}
\begin{equation}
\nu=\alpha\frac{c_s^2}{\Omega}~~~,
\end{equation}
where $\alpha$ is the viscosity parameter and $c_s$ is the sound speed at mid plane. In a steady state, mass conservation produces a surface density of
\begin{equation}
\Sigma=\frac{\dot M}{3\pi\nu}~~,
\end{equation}
where $\dot{M}$ is the infall (onto the disk) accretion rate. The surface temperature, $T_e$ in such a disk is given by energy conservation
\begin{equation}
\sigma T_e^4= \frac{9}{8}
\frac{\dot{M}}{3\pi} 
\Omega^2 + \sigma T_\textit{irr}^4~~,
\end{equation}
where $T_\textit{irr}$ is the irradiation (by the central star) temperature (e.g., \citep{Cannizzo1993}), given by 
\begin{equation}
T_\textit{irr}= \left(\frac{2}{3\pi}\right)^{1/4}
\left(\frac{R}{R_*}\right)^{-3/4} T_*~~.
\end{equation}
Here $R_*$ and $T_*$ are the star's radius and temperature~\citep{C+G1997}. The mid-plane temperature of the disk is related to its surface temperature through $T_c^4=\tau T_e^4$, where $\tau$ is the optical depth given by
\begin{equation}
\tau=
\frac{3}{8} 
\kappa 
\frac{\Sigma}{2}
\end{equation}
and the opacity is $\kappa=a T_c^b$.

\begin{figure}
	\centering
	\includegraphics[width=4.5in]{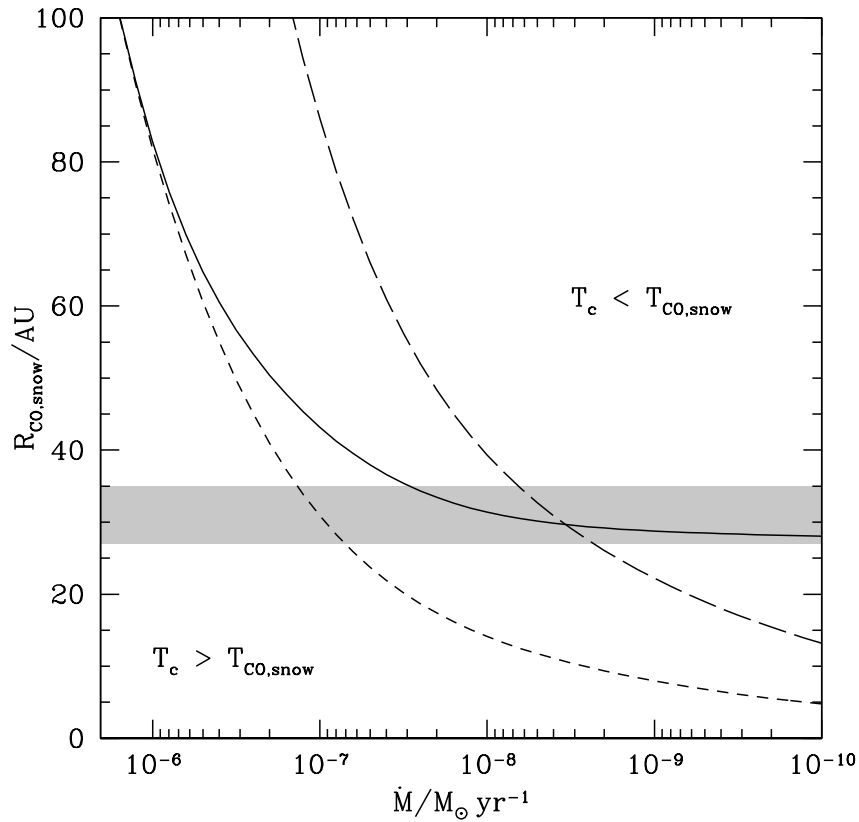}
	\caption{Evolution of the CO snow line as a function of the accretion rate in a steady state disk. The dashed lines show a fully MRI turbulent disk with a viscosity parameter of 0.01 (short-dashed) and 0.0001 (long-dashed). The solid line represents a disk with a self-gravitating dead zone. The shaded region indicates the location of the CO snow line in our solar system at the time of planetesimal formation.\label{f24}}
\end{figure}

While the precise values of $a$ and $b$ do not have a strong effect on the inferred disk's temperature, values appropriate for the low temperatures in the vicinity of the CO snow line are those obtained from absorption by dust, $a=0.053$, $b=0.74$ (e.g., \citep{Zhuetal2009}). \citet{M+L2014} solved for the CO snow line radius by equating the central temperature $T_c$ to $T_\mathrm{CO, snow}$. Their results for $R_\mathrm{CO, snow}$ as a function of the accretion rate are presented by the short-dashed curve in Figure~\ref{f24}, where the assumed parameters were $M_*=1~M_\odot$, $R_*=3~R_\odot$, $T_*=4000$~K, $T_\mathrm{CO, snow}=17$~K, and $\alpha=0.01$. Since the accretion rate drops in time, time is the implicit coordinate in this figure. The value of the viscosity parameter $\alpha$ is rather uncertain \citep{Kingetal2007}, consequently the figure also presents the results (by the long-dashed curve) obtained for $\alpha=10^{-4}$. 

The figure demonstrates that the observed location of the CO snow line in our solar system (at 27--35~AU), and in the solar system analog TW~Hydra (at 28--31~AU), \textit{cannot} be explained by a fully turbulent disk model. Specifically, in this model the CO snow line moves in too close to the central star during the low-accretion-rate phase toward the end of the disk's lifetime.

I should note that since irradiation is the dominant heating source (over viscous heating) on the scale of tens of AU, we can ignore the viscous heating term in eq.~(12), to find an approximate analytic solution that represents very closely the dashed lines in Figure~\ref{f24}, for accretion rates of up to about $10^{-8}~M_\odot$~yr$^{-1}$. This analytic approximation is given by
\begin{eqnarray}
R_\mathrm{CO, snow}^\mathrm{MRI} &\simeq
&13.2\left(\frac{\alpha}{0.01}\right)^{-2/9}
\left(\frac{M_*}{M_\odot}\right)^{1/9}
\left(\frac{\dot{M}}{10^{-8}~{M_{\odot}~\mathrm{yr}^{-1}}}\right)^{2/9}\\ \nonumber
&&\left(\frac{T_\mathrm{CO,snow}}{17~\mathrm{K}}\right)^{-0.95}
\left( \frac{R_*}{3~R_{\odot}} \right)^{2/3}
\left(\frac{T_*}{4000~\mathrm{K}}\right)^{8/9}~\mathrm{AU}~~.
\end{eqnarray}

Since the fully turbulent disk model fails to reproduce the observed CO snow line, \citet{M+L2014} also calculated time-dependent disk models with a dead zone, to follow the evolution of the CO snow line.

\subsubsection{A Disk with a Dead Zone\label{sec543}}
A dead zone is formed when the ionization fraction is not high enough for the MRI to drive turbulence. Typically, the hot part of the disk (close to the central star), where the mid-plane temperature is higher than some critical value, $T_c\simeq800$~K~\citep{U+N1988}, is thermally ionized (and therefore MRI turbulent.) Farther away from the central star, cosmic rays or the x-ray flux from the star are the dominant sources of ionization \citep{Glassgoldetal2004}. These sources can only penetrate the surface layers, down to a surface density of $\Sigma_\mathrm{crit}$. Where $\Sigma > \Sigma_\mathrm{crit}$, a dead zone exists at the mid-plane, with a surface density of $\Sigma_g=\Sigma - \Sigma_\mathrm{crit}$. The precise value of $\Sigma_\mathrm{crit}$ depends on the ionizing source, being around $\Sigma_\mathrm{crit}\simeq200$~g cm$^{-2}$ if cosmic rays dominate~\citep{Fromangetal2002}, but much lower if \mbox{x-rays} dominate~\citep{M+P2003}. In the very outer parts of the disk (in terms of distance from the central star), where $\Sigma<\Sigma_\mathrm{crit}$, the external ionization sources can penetrate all the way to the mid-plane and that part is again MRI active.

In the dead zone (low-viscosity) region, material can accumulate to the point where the disk becomes self-gravitating. This is expected to occur when the \citet{Toomre1964} parameter $Q=c_s\Omega/\pi G\Sigma$ drops below a critical value $Q_\mathrm{crit}$. \citet{M+L2014} took $Q_\mathrm{crit}=2$. This drives gravitational turbulence, with an effective viscosity parameter
\begin{equation}
\nu=\alpha_g\frac{c_s^2}{\Omega}~~.
\end{equation}
\citet{M+L2014} adopted the functional form $\alpha_g=\alpha\exp(-Q^4)$ \citep{Zhuetal2010,M+Lubow2013}.

They considered a model in which a molecular cloud collapses onto a disk, and they took the initial accretion rate to be $2\times10^{-6}~M_\odot$~yr$^{-1}$, and assumed that the accretion rate decreases exponentially, on a timescale of $10^5$ years. The initial surface density of the disk was taken to be that of a turbulent steady disk around a 1~$M_\odot$ star, with an accretion rate of $2\times10^{-6}~M_\odot$~yr$^{-1}$. The disk was modeled on a radial grid of 200 points evenly distributed in $\log R$ from $R=1$~AU to $R=200$~AU. The infalling material was added at $R=195$~AU. To allow for a comparison, \citet{M+L2014} modeled one disk to be fully turbulent, and the other to contain a dead zone, with $\Sigma_\mathrm{crit}=10$~g~cm$^{-2}$ (corresponding to ionization by x-rays).

\begin{figure}
	\centering
	\includegraphics[width=4.5in]{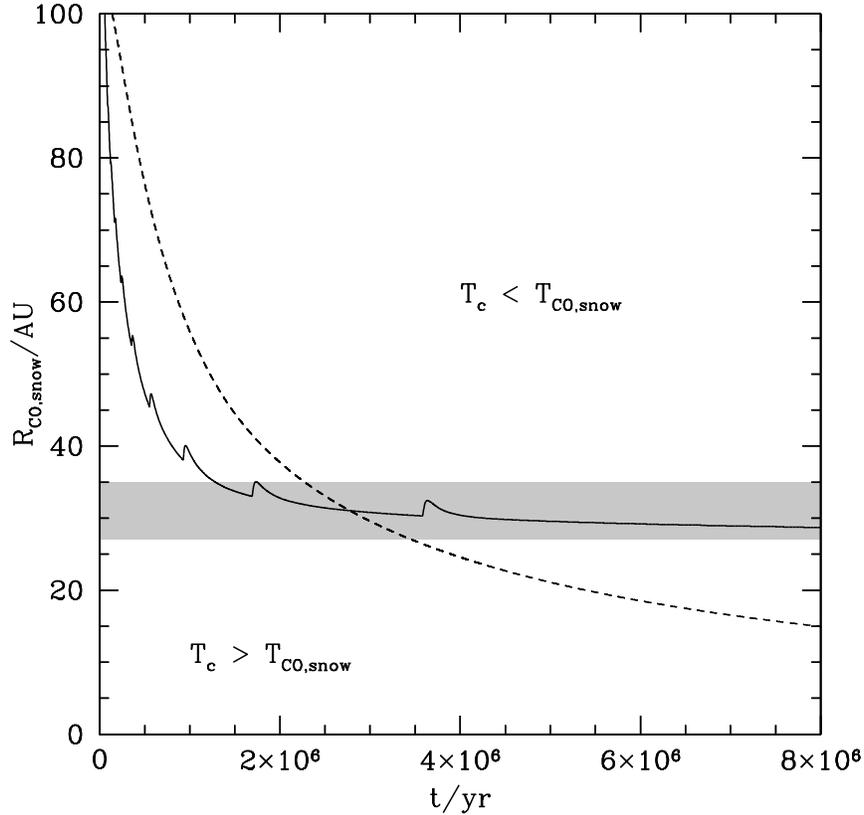}
	\caption{Evolution of the CO snow line in a time-dependent disk with an exponentially decreasing infall accretion rate. The dashed line shows a fully MRI turbulent disk and the solid line a disk with a dead zone. The shaded region indicates the location of the CO snow line in our solar system at the time of planetesimal formation.\label{f25}}
\end{figure}

The disk was found to be gravo-magneto unstable for high accretion rates, which caused unsteady accretion onto the central star (for a similar behavior see also \cite{Armitageetal2001,M+Lubow2013}. The evolution of the CO snow line as a function of time is shown in Figure~\ref{f25}. The dashed line represents the fully turbulent disk and the solid line the disk with a dead zone. At late times, the model with the dead zone has a snow line radius that is considerably larger than that obtained for a turbulent disk, and one which is consistent with the observations of the solar system and of TW~Hya. The small (and brief) repetitive increases in the snow line radius at early times are caused by FU~Orionis-type outbursts. Basically, with a dead zone, the small amount of self-gravity heats the more massive disk with a dead zone, and this causes the CO snow line radius to move outward, as required by observations. For low accretion rates, \citet{M+L2014} were able to find an analytic solution for the CO snow line radius in a disk with a dead zone. This is given by 
\begin{equation}
R_\mathrm{CO, snow}^{\rm dead}\simeq 29.3
\left(\frac{M_*}{M_\odot}\right)^{1/9}
\left(\frac{R_*}{3~R_\odot}\right)^{2/3}
\left(\frac{T_*}{4000~\mathrm K}\right)^{8/9}
\left(\frac{T_\mathrm{CO, snow}}{17~\mathrm K}\right)^{-0.61}~\mathrm{AU}~~.
\end{equation}
This approximate solution agrees very well with the solid line in Figure~\ref{f24}, where the CO snow line radius is shown as a function of the accretion rate.

I should note that the CO snow line in the disk around the Herbig Ae~star HD~163296 has been found to lie at a radius of 155~AU~\citep{Qietal2011,Mathewsetal2013}. For the observed parameters of this system: $M_*=2.3~M_\odot$, $R_*=2~R_\odot$, $T_*=9333$~K, and $\dot{M}=7.6\times10^{-8}~M_\odot$~yr$^{-1}$, the model of \citet{M+L2014} with a dead zone would predict a snow line radius of only 62~AU (a fully turbulent disk would give the even much smaller 37~AU). Disk flaring in this system could potentially account for this discrepancy, since approximations for the temperatures of flared disks would result in CO snow line radii larger than 100~AU (e.g., \citep{C+G1997}).

In general, the analytic solution for the radius of the CO snow line could prove useful in determining the composition (and thereby maybe the habitability) of exosolar planets.

\subsection{Conclusion of Sections 1--5\label{sec55}}

An examination of the physical properties of our solar system reveals that it is not extremely unusual when those are compared to the characteristics of the other observed exoplanetary systems. Still, there is no doubt that a few of the solar system's parameters have made it conducive to the emergence and evolution of life. For example, low eccentricity planets (as observed in the solar system) have a more stable temperature throughout the entire orbit, which may make them more likely to harbor life~\citep{W+P2002}. Planetary systems with a low mean eccentricity are also more likely to have a long-term dynamical stability.

The age of the solar system (about 4.5~billion years) may also be favorable for the emergence of complex life (which on Earth took some three billion years). However, I should note that since the current age of the Sun is about half of its total lifetime, and about half the age of the Milky Way's disk, one can expect that roughly half of the stars in the disk are even older than the Sun. It is still the case though, that if life \textit{can} emerge and evolve around low-mass stars (that are much more numerous and live much longer), then one could expect complex life to be much more abundant in the future (about a trillion years from now; see e.g., Chapter by A.~Loeb in this volume). If that is indeed the case, then the appearance of complex life on Earth could be regarded as very early. More recent work suggests, however, that planets around most M-dwarfs may experience serious atmosphere erosion, and are therefore not likely to harbor life~\citep{L+L2017}.

As I have already noted, the existence of terrestrial planets in the habitable zone around their host star is quite common~\citep{D+C2015,Bonfilsetal2013}.

The metallicity of the host star (and of the protoplanetary disk) also does not appear too special in the solar system. The metallicity plays a role in determining the structure of the planetary system that forms~\citep{Buchhaveetal2014,Wang+F2015}. The metallicity was found to be correlated with the probability for the star to have a giant planet orbiting it~\citep{F+V2005,Sousaetal2011,Reffertetal2015}. However, the correlation is much less clear for lower-mass planets~\citep{Buchhaveetal2012,Mayoretal2014}. While planets with radii smaller than four Earth radii are observed around stars with a wide range of metallicities, the average metallicity of stars hosting planets with $R_p<1.7~R_{\oplus}$ is close to solar~\citep{Buchhaveetal2014}. It is unclear, therefore, whether having a near-solar metallicity is somehow related to forming Earth-size planets.

The variability of the Sun has also been compared to the activity of the stars in the Kepler sample, with rather ambiguous conclusions. While Basri and collaborators~\citep{Basrietal2010,Basrietal2013} found the Sun to be quite typical, with only a quarter to a third in the sample being more active, McQuillan and his collaborators \citep{McQuillanetal2012} found the Sun to be relatively quiet, with 60\% of the stars being more active.
The difference in the conclusions stems primarily from the fact that fainter stars were included in \citet{McQuillanetal2012} (and also from some differences in defining the activity level of the Sun). Since 90\% of M~dwarfs are more active than the Sun, the inclusion of lower-mass stars makes the Sun quiet in comparison. Compared to other Sun-like stars, however, the Sun is quite typical.

As I have noted throughout Section~\ref{sec5}, the existence of a compact asteroid belt may be conducive to initiating life. Current observations do not allow us to determine categorically how common asteroid belts are. However, the list of unresolved debris disk candidates now contains hundreds of examples~\citep{Chenetal2014}, and of these about two-thirds are better modeled by a two-component (rather than a single) dust disk arising from two separate belts~\citep{K+W2014}. The two belt configuration of our solar system may therefore be quite ubiquitous.

In Sections~\ref{sec2} and~\ref{sec3} I identified the lack of super-Earths and the lack of planets interior to Mercury's orbit as perhaps the two main characteristics that make the solar system somewhat special. Even though I have shown that the presence of super-Earths could affect the rate of asteroid impacts on Earth, it is not obvious that this influenced the Earth's habitability. One feature that should be further investigated is the effect of a close-in super-Earth on the dynamical stability of a terrestrial planet in the habitable zone.

The bottom line from the discussion so far is simple. There may be many factors that are necessary, but maybe not sufficient, for life to emerge and evolve on a planet. If we were to multiply the probabilities for all of these factors (in a Drake-type equation) together, we could end up with a very small probability for life in the Milky Way. Without a concrete knowledge of which ones of these factors  are truly essential for life, however, such an exercise would merely represent our ignorance. If we blindly consider every single aspect of the solar system, we will obviously find it to be unique. From the parameters that I have considered here, however, I have not identified any feature that would argue for the Earth being exceptionally rare. This is what makes Section~\ref{sec6} particularly interesting, since in it I discuss a claim by astrophysicist Brandon Carter, who argues that extraterrestrial intelligent life is exceedingly rare.

\section{How Rare Are Extra-Solar Intelligent Civilizations?\label{sec6}}
	
\subsection{Introduction\label{sec61}}

The existence of the so-called ``Fermi Paradox,'' the absence of any signs for the existence of other intelligent civilizations in the Milky Way galaxy, coupled with an interesting argument raised by astrophysicist Brandon Carter (\citep{Carter1983} to be discussed below) as well as the absence of any physical law that mandates complexification, have convinced a few researchers that intelligent life may be exceedingly rare in the Milky Way (e.g., \citep{B+T1986,Davies2010,W+B2000}). If true, such a reality could have implications far beyond the practical ones (i.e., the search for extraterrestrial intelligence). In fact, it would fly in the face of Copernican modesty, which argues that humanity should not be in any way ``special'' in the grand cosmic scheme. Over the years, Carter's argument has generated a considerable amount of discussion, so I will briefly review it here, together with a few of the criticisms that have been raised.

\subsection{Carter's Argument\label{sec62}}

Carter's argument [138] can be explained in very simple terms as follows. Examine the typical timescale for biological evolution (and the emergence of intelligent life) on a planet, $\tau_{\ell}$, and the lifetime of the central star of that planetary system $\tau_{\ast}$. If the two timescales are \textit{a~priori} entirely independent quantities (that is, intelligent life can develop at some random time with respect to the main-sequence lifetime of the star), then one can expect that one of the following relations holds: either $\tau_{\ell}\gg\tau_{\ast}$ or $\tau_{\ell}\ll\tau_{\ast}$. The probability that $\tau_{\ell}\sim\tau_{\ast}$ is very small for two truly independent quantities, when each one of the two can assume a very broad range of values. If, however, \textit{generally} $\tau_{\ell}\ll\tau_{\ast}$, it is very difficult to understand why in the very first system in which we found an existing intelligent civilization (the Earth-Sun system), we find that $\tau_{\ell}\sim\tau_{\ast}$ (to within a factor of two; the lifetime of our Sun is about 10~billion years and it took about 4.5 billion years for intelligent life to evolve). I take $\tau_{\ell}$ to mean roughly the timescale for the appearance of land life. This means, Carter argues, that \textit{generally} $\tau_{\ell}\gg\tau_{\ast}$. In that case, however, it is clear that because of observational selection, the first system found to harbor intelligent life is likely to exhibit $\tau_{\ell}\sim\tau_{\ast}$, since a civilization would not have developed for $\tau_{\ell}\gg\tau_{\ast}$ (the evolution of life requires the star as an energy source). Consequently, Carter concludes, typically $\tau_{\ell}\gg\tau_{\ast}$, and intelligent civilizations do not develop. The Earth in this case is an extremely rare exception.

Carter's argument, if true, has very significant implications from both scientific and philosophical perspectives. In particular, it puts a very heavy burden of responsibility on humanity, as one of the very few (or only!) intelligent civilization in the Galaxy. It should come as no surprise, therefore, that several criticisms of the argument have emerged over the years since its publication.

\subsection{Criticisms of Carter's Argument\label{sec63}}

The first significant criticism of Carter's argument, primarily on logical and methodological grounds, was launched by \citet{Wilson1994}. Wilson first explained that Carter's argument really involves three timescales: $\tau_{\ast}$---the main-sequence lifetime of stars like the Sun ($\sim$10~Gyr), $\tau_{\ell}$---the time that biological evolution has taken on Earth ($\sim$4~Gyr), and $\bar{\tau}$---the timescale that would intrinsically be the most likely one required for the evolution of intelligent life. Wilson then points out that we really have no idea what the value of $\bar{\tau}$ is, since we do not understand all the biological processes involved in the appearance of intelligence, and we have only a single observed case of biological evolution (on Earth). Wilson explains that Carter's argument states that $\bar{\tau}$ has to satisfy $\bar{\tau}\ll\tau_{\ast}$ or $\bar{\tau}\gg\tau_{\ast}$, and he rules out $\bar{\tau}\sim\tau_{\ast}$ as explained above, even though we are completely ignorant about $\bar{\tau}$.

In particular, Wilson points out that when Carter assumes only the three ranges of values $\bar{\tau}\gg\tau_{\ast}$, $\bar{\tau}\ll\tau_{\ast}$, and $\bar{\tau}\sim\tau_{\ast}$, he excludes the possibility, for instance, that $\bar{\tau}$ is less than $\tau_{\ast}$, but \textit{not much less}. Wilson also argues that since we are ignorant about the value of $\bar{\tau}$, the possibility $\bar{\tau}\sim\tau_{\ast}$ cannot truly be ruled out. In addition, Wilson explains that the claim that $\tau_{\ell}$ should not differ from a given value of $\bar{\tau}$ is not equivalent to the claim that $\bar{\tau}$ should not be different from a given value of $\tau_{\ell}$, since our knowledge of the value of a statistical quantity cannot be significantly enhanced by the evidence provided by a single case. 

Finally, Wilson argues that the imprecision of the coincidence between $\tau_{\ell}$ and $\tau_{\ast}$ (a factor of two), also lessens the explanatory power of a very high value of $\bar{\tau}$, and thereby decreases the confirmatory power of the rough coincidence $\tau_{\ell}\sim\tau_{\ast}$.

Taking a different approach, I attempted to refute Carter's argument on astrophysical grounds in \citet{Livio1999}. In that paper, I used a very simple toy model for the evolution of the Earth's atmosphere to demonstrate that Carter \textit{could} be wrong, because of a fundamental assumption in his argument. Specifically, I showed that the two timescales $\tau_{\ell}$ and $\tau_{\ast}$ could, in principle at least, be correlated, in which case the entire logical structure of the argument (which is based on them being independent) collapses. Here is a brief description of how a $\tau_{\ell}$--$\tau_{\ast}$ relation \textit{could} arise.

On the face of it, it appears that $\tau_{\ell}$ is determined entirely by biochemical reactions and Darwinian evolution, while $\tau_{\ast}$ is determined by the rate  energy is produced by nuclear burning reactions and therefore, that the two timescales are indeed totally independent. However, we should note the following: The absorption of UV radiation by nucleic acids peaks in the range 2600--2700~\AA, and by proteins in the range 2700--2900~\AA\ \citep{Casperson1950,Davidson1960,Sagan1961}. Such radiation essentially kills all cell activity~\citep{Berkner1952}. The only atmospheric constituent that efficiently absorbs radiation in the 2000--3000~\AA\ range is O$_3$~\citep{Watanabeetal1958}. The appearance of land life, therefore, may have to await the accumulation of a protective layer of ozone in the atmosphere (e.g., \citep{B+M1965,Lyonsetal2014}).

The evolution of the concentration of oxygen in a planetary atmosphere is very complex, and calibrations have to rely solely on the one existing example---Earth. On Earth, dynamic rising and falling of oxygen levels started perhaps as early as 3~billion years ago, on a background of generally rising levels from low to intermediate to high (see, e.g., \citep{Lyonsetal2014} for a recent review). Most of the oxygen produced on Earth was of biotic origin. In the very early (non-biotic) stages, however (which may have lasted more than a billion years), oxygen was primarily released from the photodissociation of water vapor~\citep{Canutoetal1983}. From our perspective here, the important point is that the initial (albeit very small) rise in oxygen can be produced by nonbiological processes (in addition to the dissociation of water vapor, the splitting of CO$_2$ by intense UV radiation can also contribute; e.g., \cite{Domagaletal2014}). The duration of this phase is roughly inversely proportional to the intensity of the UV radiation in the range 1000--2000~\AA\ (since water has significant absorption peaks in the ranges 1100--1300~\AA\ and 1600--1800~\AA). Consequently, for a given planet size and orbit, the timescale for the build-up of sufficient shielding from lethal radiation (and concomitantly the timescale for the appearance of land life, $\tau_{\ell}$) is dependent on the spectral type of the star, and thereby on $\tau_{\ast}$ [since on the main sequence the spectral type is directly related to the mass, and $\tau_{\ast}\sim10^{10}(M/M_\odot)^{-2.45}$~yr]. As an aside I should note that the mere concept of a ``habitable zone'' for the planet already introduces a correlation between the star's properties and the habitability of a planet.

\citet{L+K1990} used typical main sequence relations, 
$(L/L_{\odot})=(M/M_{\odot})^{3.45}$, $(R/R_{\odot})=(M/M_{\odot})^{\beta}$ (with $\beta$ in the range 0.6--1, applicable at least for spectral types around that of the Sun or smaller), and empirical fractions for the UV radiation emitted in the 1000--2000~\AA\ range \citep{Stecker1970,Carruthers1971}, to obtain the approximate relation 
\begin{equation}
\tau_{\ell}/\tau_{\ast}\sim 0.4(\tau_{\ast}/\tau_{\odot})^{1.7}~~.
\label{eq.tau}
\end{equation}
As I have noted above, the potential existence of such a relation undermines a key assumption in Carter's argument. The fact that we found a relation of the form $\tau_{\ell}/\tau_{\ast}=f(\tau_{\ast})$, where $f(\tau_{\ast})$ is a monotonically increasing function (at least for a certain range of stellar spectral types), has another interesting consequence. For a Salpeter initial mass function~\citep{Salpeter1955}, the distribution of stellar lifetimes behaves approximately as $\psi(\tau_{\ast})\sim\tau_{\ast}$. This expresses the known fact that the number of stars increases with increasing $\tau_{\ast}$. Since $f(\tau_{\ast})$ also increases with $\tau_{\ast}$ (eq.~\ref{eq.tau}), but complex life cannot emerge if $\tau_{\ell}>\tau_{\ast}$, this implies that \textit{it is most probable that in the first place where we would encounter intelligent life we will find that $\tau_{\ell}/\tau_{\ast}\sim1$}, as in the Earth--Sun system. In other words, the observation on which Carter based his argument finds a natural explanation, and it does not have any implications for the frequency of extrasolar life. Note that this conclusion does not depend on the precise functional form of $f(\tau_{\ast})=\tau_{\ell}/\tau_{\ast}$, as long as such a relation exists, and it is a monotonically increasing function of $\tau_{\ast}$.

Since in order to obtain the $\tau_{\ell}$--$\tau_{\ast}$ relation I had to make a few extremely simplifying assumptions, I would not claim that this completely refutes Carter's argument. In particular, if instead of the one-to-one function in equation~(\ref{eq.tau}) there is an extremely wide ``band,'' this would be almost equivalent to there being no $\tau_{\ell}$--$\tau_{\ast}$ relation at all, in which case Carter's argument can be recovered. The above discussion does demonstrate, however, that Carter's argument \textit{could} be wrong from an astrophysical perspective. I should emphasize that even a complete refutation of Carter's argument does not mean that extrasolar intelligent civilizations exist---only that an argument for their nonexistence is wrong.

A criticism on a more fundamental level was raised by \citet{CVD2009}. These authors pointed out that Carter's argument relies first of all on the assumption that \textit{well-defined} timescales for the astrophysical and biological processes actually exist. Secondly, Carter assumes that the timescale on the astrophysical side is even fixed, and approximately known. \citet{CVD2009} noted in particular that processes other than the evolution of the host star can affect the habitability of a planet. They correctly asserted that Carter's argument regards the Earth--Sun system as a ``closed box,'' while concepts such as the Galactic habitable zone (the region in the Galaxy characterized by such physical parameters that it allows for life to emerge and evolve; e.g., \cite{Gonzalez2005,Lineweaveretal2004}) demonstrate a level of connectedness not envisaged by Carter. Other effects such as ``snowball Earth,'' glaciation episodes, and geophysical processes such as those governing the carbon-silicate cycle also demonstrate the existence of relevant timescales shorter than the main-sequence lifetime of the host star (e.g., \citep{Hoffmanetal1998,G+Y2003}). I should also point out that since some astrophysical processes and events exist (such as Gamma-ray bursts [GRBs]; or spiral arms crossings by the solar system), that can altogether terminate or at least strongly affect biological evolution (see e.g., \citep{Piranetal2016,Sloanetal2017} for recent references on GRBs and \cite{Shaviv2002} on spiral arms crossings), the governing timescales may be the ones associated with maximizing the chances of life being able to survive these cosmic cataclysms, rather than the lifetime of the star (the timescale for atmospheric loss for planets around M-dwarfs is also much shorter than the main-sequence lifetime~\citep{L+L2017}).  These timescales are dictated primarily by the values of the Hubble constant H$_\mathrm{o}$, and the value of the cosmological constant $\Lambda$ (in the context of a $\Lambda$CDM cosmology). It is interesting to note in this respect that using cosmological $N$-body simulations, \citet{Piranetal2016} concluded that we find ourselves in a favorable spot in the cosmological phase space, in that the exposure of the solar system to GRBs is minimized, while the number of hydrogen-burning stars (around which complex life can in principle evolve) is maximized.

The bottom line is clear: Carter's argument should not be taken as a discouragement from searching for other intelligent civilizations in the Milky Way. With the realization that punctuated equilibrium could characterize the evolution of life even on the Galactic scale, even the possibility of $\tau_{\ell}\ll\tau_{\ast}$ cannot be convincingly rejected, since life can independently appear several times and then become extinct through catastrophic events.

Irrespective of Carter's argument, however, there are other reasons to suspect that we are not the only intelligent species to have ever existed in the universe as a whole. For example, \citet{F+S2016} used the exoplanets statistics obtained by the Kepler Space Observatory to evaluate the probability that humanity is the only technological civilization to have ever existed. They showed that for that to be true, the probability that a habitable-zone planet develops a technological species must be smaller than $10^{-24}$, essentially the reciprocal of the expected number of rocky, habitable-zone planets in the observable universe. So, unless the evolution to technology is truly  extraordinarily improbable, chances are that other such civilizations existed at some point in the universe's lifetime.

To conclude this entire chapter, I have not identified any physical parameters which convincingly demonstrate that life on Earth, or even intelligent life, is unique, either in the universe as a whole, or even in the Milky Way galaxy. Upcoming space missions such as TESS and JWST (to be launched in 2018 and 2019, respectively) and WFIRST (to be launched in the mid-2020s), as well as ground-based telescopes (such as a next generation Extremely Large Telescope) will detect and start to characterize the atmospheres of super-Earth and Earth-like planets, in the search for biosignatures. Chances are that even if we do not detect extrasolar life in the next 2--3~decades, we will at least be able to place some meaningful limits on how rare life that dominates the planetary surface chemistry (so that it significantly alters the atmosphere) really is.

\section*{Acknowledgements}
Most of the work presented in this chapter was done in collaboration with Rebecca Martin. I am also grateful to Jim Pringle, Andrew King, Joe Silk, Jeremy Smallwood, Stephen Lepp, Martin Beer, Steve Lubow, Phil Armitage, Arik Kopelman, and Lev Yungelson, who contributed to various parts of the work.

\end{document}